\documentclass[10pt,journal,twocolumn]{IEEEtran}
\IEEEoverridecommandlockouts
\usepackage{cite}
\usepackage{amsmath,amssymb,amsfonts,booktabs}
\usepackage{graphicx}
\usepackage{textcomp}
\usepackage{xcolor}
\usepackage{enumerate}
\usepackage{epstopdf}
\usepackage{subcaption}
\usepackage{amsthm}
\newtheorem{theorem}{Theorem}

\usepackage{algorithm}
 \usepackage{algpseudocode}
 \newcommand{\revise}{\textcolor{black}}
\allowdisplaybreaks[4]

\newtheorem{remark}{Remark}
\newtheorem{lemma}{Lemma}

\def\BibTeX{{\rm B\kern-.05em{\sc i\kern-.025em b}\kern-.08em
		T\kern-.1667em\lower.7ex\hbox{E}\kern-.125emX}}

\begin{document}

\title{LAWNs Meet SWIPT: Beamforming and Power Splitting Optimization for Predictive Control 
}

  \author{
    \IEEEauthorblockN{Jun Wu, \IEEEmembership{Graduate Student Member, IEEE,}  Wenchao Liu, \IEEEmembership{Graduate Student Member, IEEE}, Weijie Yuan, \IEEEmembership{Senior Member, IEEE}, Nanchi Su, \IEEEmembership{Member, IEEE}
  }
  \thanks{
  This work of Weijie Yuan was supported in part by the National Natural Science Foundation of China under Grants 62471208 and U25A20389; in part by the Guangdong Provincial Natural Science Foundation under Grant 2026B1515020041; in part by the Shenzhen Science and Technology Program under Grant JCYJ20240813094627037. This work of Nanchi Su was supported in part by the National Natural
Science Foundation of China under Grants 62401181. (\textit{Corresponding author: Weijie Yuan})
  
  J. Wu, W. Liu, and W. Yuan  are with the School of Automation and Intelligent Manufacturing, Southern University of Science and
  Technology, Shenzhen 518055, China (email: wuj2021@mail.sustech.edu.cn; wc.liu@foxmail.com; yuanwj@sustech.edu.cn )
  
  N. Su is with the Guangdong Provincial Key Laboratory of Aerospace Communication and Networking Technology, Harbin Institute of Technology (Shenzhen), Shenzhen 518055, China (e-mail: sunanchi@hit.edu.cn).
  }
  }

\maketitle

\begin{abstract}
Simultaneous wireless information and power transfer (SWIPT) has emerged as a promising paradigm for enabling sustainable connectivity in battery-limited low-altitude wireless networks (LAWNs). This paper investigates a SWIPT-enabled LAWN system in which a multi-antenna base station (BS) simultaneously delivers control information and wireless energy to a fleet of uncrewed aircraft systems (UASs) via power splitting. In particular, the BS remotely guides the UASs to accurately track predefined reference trajectories toward their destinations while avoiding multiple mobile no-fly zones (NFZs). To guarantee collision-free path planning, we first construct smooth and safe reference trajectories using stream function theory. Then, a real-time optimization problem is formulated, which jointly takes into account the wireless control cost and energy sustainability by optimizing control inputs, transmit beamforming vectors, and the power splitting ratios. To address the resultant non-convex problem, a two-stage optimization framework is proposed. First, we develop a model predictive control (MPC)-based method to generate predictive control inputs. Subsequently, we derive a computationally efficient iterative algorithm to optimize the beamforming vectors and power splitting ratios by applying semidefinite relaxation (SDR) and successive convex approximation (SCA) techniques. We further prove that the SDR is tight for our formulation. Extensive numerical results demonstrate that our proposed design significantly outperforms benchmark schemes in terms of tracking accuracy and harvested energy, thereby validating its effectiveness for sustainable implementation in LAWN systems.
\end{abstract}


\section{Introduction}

The emergence of sixth-generation (6G) wireless systems will extend connectivity into three-dimensional (3D) space, positioning low-altitude wireless networks (LAWNs) as a foundational component of future communication infrastructure \cite{wu2025low,yuan2025ground,11017717}. LAWNs integrate coordinated uncrewed aircraft systems (UASs) with terrestrial access networks, backhaul links, and edge computing resources to establish persistent air-to-ground (A2G) links that support both control signaling and high-throughput data transmission. 
This capability enables LAWNs to support a wide range of mission-critical and economically promising applications, such as autonomous transportation, industrial inspection, and intelligent logistics. For instance, in last-mile delivery, LAWNs can reduce end-to-end latency, thereby enhancing routing flexibility in congested urban environments and lowering operational costs by shortening travel paths. 
Consequently, LAWNs have recently attracted significant research interest from both academia and industry.

To fully unlock the potential of LAWNs, communication plays a central and enabling role, providing not only the fundamental connectivity for exchanging control commands, state information, and mission data among distributed aerial and ground units, but also supporting coordination and real-time decision-making in dynamic environments \cite{wu2025towardm}. Significant research efforts have therefore been devoted to advancing air-to-ground (A2G) and air-to-air (A2A) communication techniques, including spatial beamforming design \cite{11206496,10596930}, UAS trajectory optimization \cite{10168298,10654366}, and spectrum-efficient resource allocation \cite{cai2020joint}. However, in contrast to traditional UAS-based communication systems, where the UASs are deployed as passive relays and aerial base stations (BSs) to extend connectivity to underserved ground users, LAWNs operate under a broader functional paradigm, in which each UAS serves as an active agent capable of executing complex autonomous tasks. In this context, a fundamental prerequisite is the ability to navigate complex, shared airspace for safe and large-scale deployments. In practice, the low-altitude airspace is not fully open, which is affected by no-fly zones (NFZs) imposed by various factors, such as tall buildings, restricted military areas, and occupied civil airspaces, necessitating advanced collision-free trajectory design \cite{kahne2002air}. The authors in \cite{xu2020multiuser} investigated a robust resource allocation algorithm for multiuser downlink multiple-input single-output (MISO) UAS communication systems, where the design is formulated as a non-convex optimization problem accounting for the imperfect angle of departure (AoD), wind speed uncertainty, and
polygonal NFZs. Similarly, the work in \cite{heo2024joint} introduced a new constraint that allows the UAS to perfectly avoid NFZs throughout the entire continuous trajectory with rigorous mathematical proof. Under this framework, the scheduling, transmit power, time-slot duration, and trajectory of the UAS were studied, where the goal was to maximize the minimum throughput among ground nodes without violating NFZs. Although these studies well addressed path planning under resource constraints, the resulting formulations are typically highly non-convex, making it intractable to achieve globally optimal solutions. In large-scale LAWN deployments, the computational burden grows rapidly, rendering the aforementioned approaches inapplicable to battery-limited LAWN systems.

\revise{To reduce the computational complexity of collision-free trajectory generation, several methods have been proposed. For instance, a virtual potential field method was introduced in \cite{xiao2018low}, where target locations and obstacles are integrated into a continuously updated potential field framework.} This approach reduces the complexity of high-dimensional optimization and reduces the number of decision variables, enabling scalable 3D path generation in complex environments. Building on potential field theory, the work in \cite{1241966} developed an analytical navigation framework based on hydrodynamic stream functions to generate smooth, local-minima-free trajectories in two-dimensional (2D) planar airspace. The stream-function formulation provides exact and continuous path solutions while ensuring collision avoidance and controllable motion behavior. Despite their contributions, these methods primarily rely on static representations of the environment, assuming fixed NFZs. This assumption does not hold in LAWNs, where UASs operate in dynamic and regulated airspace subject to temporary flight restrictions, pop-up NFZs, and time-varying safety constraints. \revise{Path planning that explicitly accounts for mobile NFZ avoidance and dynamic airspace adaptation has received relatively limited attention. Moreover, most existing studies focus on path planning under simplified settings and do not adequately consider environmental uncertainties, such as wind disturbances. Consequently, the actual flight trajectory may deviate significantly from the planned path, which calls for effective control strategies to ensure accurate trajectory tracking in practical environments.}
 
Recent research has shifted from traditional trajectory planning toward control-oriented design that explicitly considers system dynamics, safety constraints, and execution reliability\cite{jin2025advancing}. For example, the authors in \cite{10716680,pan1,pan3} reformulated obstacle-aware motion planning as a nonlinear programming problem and introduced exact collision-avoidance constraints that reduce auxiliary variables, improving computational efficiency. In \cite{pan4,wang2019path,pan2}, a quadratic programming (QP) framework was developed using a safe flight corridor to enable real-time receding-horizon path planning in cluttered environments. Similarly, the work in \cite{li2023real} integrated approximate convex optimization with model predictive control (MPC) to achieve dynamically feasible, collision-free trajectories under vehicle and comfort constraints. Nevertheless, the above works are not well-suited to LAWN systems. On the one hand, they implicitly assume sufficient onboard computational resources for real-time optimization, which is impractical for resource-constrained UAS platforms. In LAWNs, control often relies on wireless feedback, where ground BSs or edge servers assist in decision-making and transmitting control signals via A2G links. Consequently, wireless communication factors, such as packet loss, channel outages, and limited spectral efficiency, directly affect closed-loop control performance. For instance, if the UAS does not receive sufficient information to decode the control signal arriving from the BS, the entire control system may suffer from destabilization. On the other hand, the high mobility of UASs makes LAWN operations highly delay-sensitive, and neglecting communication latency can severely degrade trajectory tracking performance. To mitigate this challenge, recent studies have adopted finite blocklength (FBL) transmission to support control signaling under stringent latency constraints, e.g.,\cite{liu2023joint,jiao2024uav}. Unlike Shannon-capacity-based designs, FBL characterizes the rate–reliability–latency tradeoff for short control packets, enabling timely and dependable command delivery in time-critical scenarios. For example, the work in \cite{jin2025predictive} integrates MPC with FBL transmission and jointly optimizes control policy, transmit power, and UAS trajectory. By solving the resulting problem via alternating optimization (AO), the proposed method improves tracking performance under communication constraints, as demonstrated by simulations and AirSim experiments. However, this study is limited to a single-UAS setting, and a scalable wireless control framework for multi-UAS coordination in LAWNs has not been reported in the open literature yet.
 
Furthermore, the sustainability and operational reliability of LAWNs are fundamentally constrained by the limited battery capacity of UAS platforms. Ensuring long-term operational viability is therefore essential for practical deployments. Conventionally, energy efficiency (EE) has been widely used as a key performance metric, and numerous studies, such as \cite{11159297,li20213d,abrar2021energy}, have investigated EE maximization by jointly optimizing trajectory planning and transmit power control. Although these approaches extend network lifetime, they often come at the expense of degraded system performance due to stringent energy-saving constraints. \revise{To address this limitation, energy harvesting (EH) techniques have been explored to enable UASs to replenish their energy from ambient sources, such as solar energy \cite{wei2021resource, pan2017performance, pan2016secrecy}. However, ambient energy sources are inherently intermittent and weather-dependent, making them unsuitable for providing stable and predictable power in LAWNs. As an alternative, wireless power transfer (WPT) offers a viable solution, although it is limited by significant path loss, which restricts its effectiveness to short-range charging. To enhance charging efficiency, radio-frequency (RF)-enabled WPT has been employed to power low-power devices, with multi-antenna techniques playing a crucial role in improving transfer efficiency. Furthermore, as RF signals can simultaneously transmit both information and energy, simultaneous wireless information and power transfer (SWIPT) has emerged as a promising approach to improve energy sustainability, enabling UASs to harvest energy from communication signals while maintaining connectivity.} In \cite{9136595}, a UAS-enabled SWIPT was investigated to simultaneously address energy scarcity and connectivity challenges in emergency Internet of Things (IoT) networks. In fact, SWIPT inherently exhibits a fundamental trade-off between information decoding and EH, since allocating more received power to EH reduces the signal-to-noise ratio (SINR) available for decoding, and vice versa. Consequently, extensive research has focused on characterizing this trade-off and developing advanced power-splitting (PS) schemes to achieve an optimal balance under various conditions. This trade-off becomes significantly more critical in LAWNs, where wireless links serve not only data transmission but also carry delay-sensitive control commands. In such settings, conventional joint communication–EH designs are insufficient, as they ignore the impact of communication quality on closed-loop stability and tracking performance. Nevertheless, a unified framework that jointly optimizes wireless control performance and sustainability in SWIPT-enabled LAWNs remains an open challenge.

\revise{Motivated by the above, this paper investigates the design of wireless control in SWIPT-enabled LAWN systems in the presence of multiple mobile NFZs. We consider a multi-antenna BS that simultaneously serves as both a WPT source and a communication transmitter, enabling the concurrent delivery of wireless control and energy transfer to multiple UASs via a shared RF signal. Our goal is to jointly optimize the transmit beamforming, PS ratios, and control inputs to ensure that each UAS accurately tracks a predefined, collision-free trajectory while harvesting sufficient energy for long-term operation. The main contributions of this work are as follows:}
\begin{itemize}
 \item To guarantee safe navigation in dynamic airspace, a stream function-based reference path design is developed to provide analytical and smooth guidance fields that inherently satisfy mobile NFZ constraints. The resulting trajectories ensure continuous collision avoidance with low computational complexity, enabling real-time applicability in dynamic LAWN environments.
\item We formulate a real-time optimization problem that minimizes the trajectory tracking error per harvested energy under system stability constraints and maximum available transmit power budget by jointly optimizing the transmit beamforming, the PS ratios, and the control inputs. To meet stringent delay requirements, we propose to adopt FBL transmission for time-sensitive control signals.
\item To address the resultant non-convex optimization problem, we develop a two-step optimization framework. In particular, the control inputs are obtained via a standard QP. The transmit beamforming and PS ratio are optimized via an AO method, where the semidefinite relaxation (SDR) and successive convex approximation (SCA) techniques are employed. The rank-one optimality of the proposed algorithm is analyzed.
\end{itemize}
Extensive simulations demonstrate the effectiveness of our proposed method in terms of tracking accuracy and energy sustainability compared with baseline schemes. The remainder of this paper is organized as follows. Section \ref{SEC2} presents the system model. Section \ref{SEC3} formulates the joint optimization problem. The proposed solutions are detailed in Sections \ref{SEC4} and \ref{SEC5}. Simulation results are presented in Section \ref{SEC6}, while Section \ref{SEC7} concludes this paper.

\revise{\textit{Notations:} The boldface lowercase letter and boldface capital letter denote the vector and the matrix, respectively. The superscript $(\cdot)^{\top}$ and $ (\cdot)^{\mathrm{H}}$ stand for the transposition and Hermitian operations, respectively.  $Q^{-1}(\cdot)$ is the inverse of the Gaussian Q-function $Q(x) = \frac{1}{\sqrt{2\pi}} \int_x^{\infty} \exp\left(-\frac{t^2}{2}\right) dt$. We use $\mathcal{CN} (\mu,v)$ to depict a complex Gaussian distribution of mean $\mu$ and variance $v$ and $\dot{x}$ to denote the first derivative of a continuous variable $x$.  $\mathbb{E(\cdot)}$ represents the
statistical expectation and $\mathrm{tr}(\cdot)$ represents the trace operation. $|\cdot|$ and $|| \cdot||$ indicate the absolute value and Euclidean norm, respectively.}

\section{System Model} \label{SEC2}
 In this section, we first introduce the UAS dynamics model and the corresponding control model, followed by presenting the SWIPT model.
\begin{figure}
    \centering
    \includegraphics[width=\linewidth]{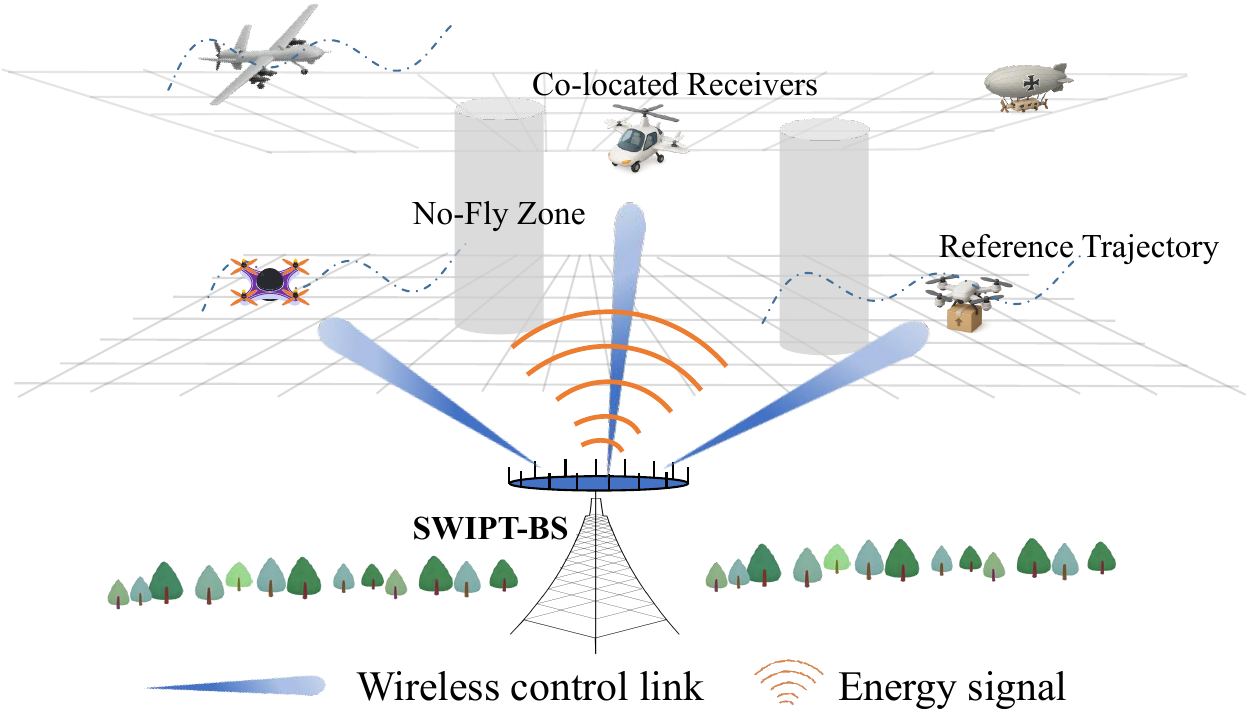}
    \caption{The considered SWIPT-LAWN scenario, where the BS simultaneously delivers control information and wireless energy to a fleet of UASs.}
    \label{fig1}
\end{figure}

\subsection{UAS Dynamics and Control Model}
\label{subsec:vehicle_dynamics}
As illustrated in Fig.~\ref{fig1}, we consider a SWIPT-enabled LAWN system, where a BS equipped with $N_t$ antennas serves $K$ single-antenna UASs operating within low-altitude airspaces.  In practice, the airspace may also contain multiple NFZs that impose additional constraints on UAS operations. To ensure safe navigation, the BS acts as a remote controller by transmitting navigation signals, while also providing power to improve the UAS sustainability. Specifically, the BS remotely controls the UASs to track a preset reference trajectory. Moreover, each UAS is assumed to be equipped with a co-located receiver (Rx) that splits the received signal into two streams in which one portion, with the PS ratio $\alpha_k,\ k \in \mathcal{K}=\{1,\ldots,K\}$, is directed to the information decoding (ID) module, and the remaining portion with ratio ($1-\alpha_k$) is allocated to the EH module.

We consider that each UAS is operating at a constant altitude $H_k$ over time. Let $\mathbf{q}_k=[x_k,y_k]^\top$ and $\theta_k$ denote the time-varying horizontal position and heading angle associated with the $k$-th UAS, respectively. Hence, the kinematic model can be simply expressed as
\begin{equation}
\left\{
\begin{aligned}
\dot x_k &= v_k\cos\theta_k,\\
\dot y_k &= v_k\sin\theta_k,\\
\dot\theta_k &= \omega_k,
\end{aligned}\right.
\label{eq:kinematic_model}
\end{equation}
where $v_k$ is the forward speed and $\omega_k$ denotes the angular velocity. In light of \eqref{eq:kinematic_model}, the second-order derivative of $\mathbf q_k$ can be straightforwardly derived by
\begin{equation}
\ddot{\mathbf q}_k =
\begin{bmatrix}
\cos\theta_k & -v_k\sin\theta_k\\[4pt]
\sin\theta_k & \;\;v_k\cos\theta_k
\end{bmatrix}
\begin{bmatrix}
\dot v_k\\[4pt]
\omega_k
\end{bmatrix}\triangleq\begin{bmatrix}
 u_{1,k}\\[4pt]
u_{2,k}
\end{bmatrix},
\label{eq:z_ddot}
\end{equation}
where $\mathbf u_k=[u_{1,k}\; u_{2,k}]^\top$ is further introduced as the control input vector in the case of $v_k\neq0$. Then, by rearranging \eqref{eq:z_ddot} and combining with \eqref{eq:kinematic_model}, we have the following state space model
\begin{equation}
\left\{
\begin{aligned}
\dot x_k &= v_k\cos\theta_k,\\
\dot y_k &= v_k\sin\theta_k,\\
\dot\theta_k &= -u_{1,k}\sin\theta_k/v_k + {u}_{2,k}{\cos\theta_k}/v_k,\\
\dot v_k &= u_{1,k}\cos\theta_k + u_{2,k}\sin\theta_k.
\end{aligned}
\right.
\label{eq:nonlinear_model}
\end{equation}
Let $\mathbf{{x}}_k=[x_k,y_k,\theta_k,v_k]^\top$ be the state vector associated with the $k$-th UAS, such that \eqref{eq:nonlinear_model} can be recast to a more compact form, i.e., $
\mathbf{\dot{x}}_k=f(\mathbf{x}_k,\mathbf{u}_k).
 $
 However, it is noted that the nonlinearity of \eqref{eq:nonlinear_model} hinders the effective control policy design. To compromise, we now linearize \eqref{eq:nonlinear_model} by adopting the first-order Taylor expansion around the local points $\left(\mathbf{\bar{x}}_k,\mathbf{\bar{u}}_k\right)$, which follows that 
    \begin{align}
   \nonumber \dot{\mathbf{x}}_k=f\left(\mathbf{\bar{x}}_k,\mathbf{\bar{u}}_k\right)+\frac{\partial f(\mathbf{x}_k,\mathbf{u}_k)}{\partial \mathbf{x}_k} \bigg|_{(\mathbf{\bar{x}}_k,\mathbf{\bar{u}}_k)} (\mathbf{x}_k-\mathbf{\bar{x}}_k)\\+\frac{\partial f(\mathbf{x}_k,\mathbf{u}_k)}{\partial \mathbf{u}_k} \bigg|_{(\mathbf{\bar{x}}_k,\mathbf{\bar{u}}_k)} (\mathbf{u}_k-\mathbf{\bar{u}}_k),
\end{align}
where { \small
\begin{align} \label{mata}
    \nonumber&\frac{\partial f(\mathbf{x}_k,\mathbf{u}_k)}{\partial \mathbf{x}_k} \bigg|_{(\mathbf{\bar{x}}_k,\mathbf{\bar{u}}_k)}
     =\\ &\left[\begin{array}{cccc}
0 & 0 & -\bar{v}_k \sin \bar{\theta}_k & \cos \bar{\theta}_k \\
0 & 0 & \bar{v}_k \cos \bar{\theta}_k & \sin \bar{\theta}_k \\
0 & 0 & \frac{-\bar{u}_{1,k} \cos \bar{\theta}_k-\bar{u}_{2,k} \sin \bar{\theta}_k}{\bar{v}_k } & \frac{\bar{u}_{1,k} \sin \bar{\theta}_k-\bar{u}_{2,k} \cos \bar{\theta}_k}{\bar{v}_k^2} \\
0 & 0 & \bar{u}_{2,k} \cos \bar{\theta}_k-\bar{u}_{1,k} \sin \bar{\theta}_k & 0
\end{array}\right],
\end{align}}
and 
\begin{align} \label{matb}
     \frac{\partial f(\mathbf{x}_k,\mathbf{u}_k)}{\partial \mathbf{u}_k} \bigg|_{(\mathbf{\bar{x}}_k,\mathbf{\bar{u}}_k)}
     = \left[\begin{array}{cc}
0 & 0 \\
0 & 0 \\
-\frac{\sin \bar{\theta}_k}{\bar{v}_k} & \frac{\cos \bar{\theta}_k}{\bar{v}_k} \\
\cos \bar{\theta}_k & \sin \bar{\theta}_k
\end{array}\right] ,
\end{align}
with $\bar{v}_k, \bar{\theta}_k \in \mathbf{\bar{x}}_k$ and $\bar{u}_{1,k}, \bar{u}_{2,k} \in \mathbf{\bar{u}}_k$, respectively. By denoting the change in the state quantity error and control input error as $\Tilde{\mathbf{x}}_k=\mathbf{x}_k-\bar{\mathbf{x}}_k$ and $\Tilde{\mathbf{u}}_k=\mathbf{u}_k-\bar{\mathbf{u}}_k$,  we then arrive
\begin{align} \label{cont}
    \dot{\Tilde{\mathbf{x}}}_k=\mathbf{A}_k{\Tilde{\mathbf{x}}}_k+\mathbf{B}_k {\Tilde{\mathbf{u}}}_k,
\end{align}
where the matrices $\mathbf{A}_k\in\mathbb{R}^{4\times4}$ and  $\mathbf{B}_k\in\mathbb{R}^{4\times2}$ are denoted in \eqref{mata} and \eqref{matb}, respectively. To facilitate the system design in the sequel, the forward Euler method is leveraged to discretize the continuous dynamics model\cite{li2023real}. By doing so, \eqref{cont} becomes that 
\begin{align}
\dot{\Tilde{\mathbf{x}}}_k=\frac{{\Tilde{\mathbf{x}}}_{k,n+1}-{\Tilde{\mathbf{x}}}_{k,n}}{\Delta t}=\mathbf{A}_k{\Tilde{\mathbf{x}}}_{k,n}+\mathbf{B}_k {\Tilde{\mathbf{u}}}_{k,n}, \label{dis}
\end{align}
in which ${\Tilde{\mathbf{x}}}_{k,n}$ and ${\Tilde{\mathbf{u}}}_{k,n}$ represent the discrete state and control input in the $n$-th time slot. The term $\Delta t$  denotes sampling duration. After some mathematical manipulations, the discrete state-space model follows that 
\begin{align} \label{spacestate}
    \left\{ \begin{array}{ll}
\Tilde{\mathbf{x}}_{k,n+1}=\underbrace{(\Delta t\mathbf{A}_k+\mathbf{I})}_{\Tilde{\mathbf{A}}_k} {\Tilde{\mathbf{x}}}_{k,n}+\underbrace{\Delta t\mathbf{B}_k}_{\tilde{\mathbf{B}}_k}{\Tilde{\mathbf{u}}}_{k,n}, \\
\mathbf{\tilde{y}}_{k,n}=\mathbf{D}_{k}\mathbf{\tilde{x}}_{k,n},
    \end{array} \right.
\end{align}
with $\mathbf{\tilde{y}}_{k,n}$ and $\mathbf{D}_k\in\mathbb{R}^{4\times4}$ being the control output and the corresponding output matrix, respectively.

\subsection{SWIPT Model}

In the $n$-th time slot, let $\mathbf{s}_n$ be the baseband transmitted signal at the BS, given by
\begin{equation}
\mathbf{s}_n = \sum_{k=1}^{K} \mathbf{w}_{k,n} c_{k,n} + \mathbf{v}_n, \label{eq4}
\end{equation}
where $c_{k,n} \sim \mathcal{CN}(0,1)$ 
denotes the modulated symbol intended for UAS $k$ in time slot $n$ 
and $\mathbf{w}_{k,n} \in \mathbb{C}^{N_t \times 1}$ 
is the corresponding control signal transmit beamforming vector. 
The vector $\mathbf{v} \in \mathbb{C}^{N_t \times 1}$ 
represents the energy signal modeled as a complex pseudo-random sequence 
with covariance matrix $\mathbf{V} = \mathbb{E}\{\mathbf{v}\mathbf{v}^H\}$. Hence, the covariance matrix of $\mathbf{s}_n$ can be straightforwardly written as $\mathbf{C}_n= \sum_{k=1}^K\mathbf{w}_{k,n}\mathbf{w}_{k,n}^H+\mathbf{V}_n$.
Let $\mathbf{h}_{k,n} \in \mathbb{C}^{N_t \times 1}$ 
be the channel vector from the BS to UAS $k$ in time slot $n$. In light of  \cite{10168298,10654366,cai2020joint}, it is assumed that the channels from the BS to UASs are dominated by line-of-sight (LoS) links, and the Doppler effects caused by UAS mobility are well compensated. Thus, the channel vector from the BS to the UAS $k$ can be expressed as 
\begin{align}
    \mathbf{h}_{k,n}=\sqrt{\frac{\beta_0}{||\mathbf{q}_{k,n}-\mathbf{g}||^2 + (H_k - H_{\rm{g}})^{2}}}\mathbf{a}_{k,n}, \label{channelgain}
\end{align}
where $\beta_0$ denotes the channel power gain at a unit reference distance and $H_{\rm{g}}$ denotes the altitude of the BS. \revise{It is assumed that perfect channel state information (CSI) is available at the transmitter to facilitate the tractable system design.} The vectors $\mathbf{q}_{k,n}$ and $\mathbf{g}$ represent the horizontal locations of UAS $k$ in time slot $n$ and the BS, respectively. 
The term $\mathbf{a}_{k,n}$ is the steering vector associated with UAS $k$, given by 
\begin{align}
  \mathbf{a}_{k,n}=\left[1, e^{j 2 \pi \frac{d}{\lambda} \cos \phi_{k,n}}, \ldots, e^{j 2 \pi \frac{d}{\lambda}(N_t-1) \cos \phi_{k,n}}\right]^{\top}, \label{steering}
\end{align}
where $d$ denotes the spacing between two adjacent antennas, $\lambda$ denotes the carrier wavelength, and $ \phi_{k,n}=\arccos \frac{H_k - H_{\rm{g}}}{\sqrt{\left\|\mathbf{q}_{k,n}-\mathbf{g}\right\|^2+ (H_k - H_{\rm{g}})^{2}}}$ denotes the AoD from the BS to the UAS $k$ at time slot $n$.
In time slot $n$, the received signal for the UAS $k$ is given by
\begin{equation}
r_{k,n} = \mathbf{h}_{k,n}^H \left( \sum_{k'=1}^{K} \mathbf{w}_{k',n} c_{k',n} + \mathbf{v}_n \right) + \omega_k, \label{eq5}
\end{equation}
where $\omega_k \sim \mathcal{CN}(0,\sigma_k^2)$ 
is the AWGN at the receiver with variance $\sigma_k^2$. Consequently, the received RF power at UAS $k$ can be derived by 
\begin{equation}
P_{k,n}^{\mathrm{r}}= \mathbb{E}\{|r_{k,n}|^2\} = 
\sum_{k'=1}^{K} \mathrm{tr}(\mathbf{W}_{k',n} \mathbf{H}_{k,n}) + \mathrm{tr}(\mathbf{V}_n \mathbf{H}_{k,n}) + \sigma_k^2, \label{eq6}
\end{equation}
where $\mathbf{W}_{k',n} = \mathbf{w}_{k',n}\mathbf{w}_{k',n}^H$ 
and $\mathbf{H}_{k,n}= \mathbf{h}_{k, n} \mathbf{h}_{k,n}^H$. 
Notice that the received power at each UAS can be allocated for multiple purposes. A portion of the received power is used for decoding the intended information signal, while the remaining fraction can be harvested for prolonging lifetime.
In particular, UAS $k$ decodes the corresponding data stream by leveraging the split received power with ratio $\alpha_{k,n}$, such that the received SINR is given by 
\begin{align}
&\mathrm{SINR}_{k,n} \nonumber \\&=\frac{\alpha_{k,n} \, \mathrm{tr}(\mathbf{W}_{k,n} \mathbf{H}_{k,n})} {\alpha_{k,n}\left(\mathrm{tr}\!\left(\sum_{k' \neq k} \mathbf{W}_{k',n} \mathbf{H}_{k,n} + \mathbf{V}_n\mathbf{H}_{k,n} \right)+ \sigma_k^2\right)+ \sigma_p^2},
\end{align}
with $\sigma_p^2$ being the processing noise power 
and the term $\sum_{k' \neq k} \mathbf{W}_{k',n} \mathbf{H}_{k,n}$ representing the multi-user interference (MUI). Thus, the achievable spectral efficiency (SE) is expressed as 
\begin{equation}
R_{k,n} = \log_2 \!\left( 1 +\mathrm{SINR}_{k,n} 
\right), \label{eq9}
\end{equation}
On the other hand, the UAS $k$ allocates the remaining received power at the analog RF front-end with a PS ratio $(1-\alpha_k)$ for EH. By adopting a linear EH model, the harvested power at UAS $k$ follows that
\begin{equation}
P^{\text{e}}_{k,n} = \eta_k(1-\alpha_{k,n}) P_{k,n}^{\mathrm{r}},
\end{equation} 
where $0<\eta_k\leq 1$ denotes the constant energy conversion efficiency\footnote{\revise{In this study, a linear EH model is adopted to facilitate the tractability of the joint predictive control and beamforming optimization. The investigation of more complex nonlinear EH models is deferred to our future research.}}. 

\section{Problem Formulation}\label{SEC3}
To ensure that the UASs closely follow the reference trajectory, a MPC-based trajectory tracking method is employed, which resorts to a receding-horizon scheme that predicts the UAS dynamics over a finite horizon and penalizes the deviation from a time-varying reference. As a result, the control cost function of the UAS $k$ is expressed as \cite{jin2025predictive}
\begin{align}
  \nonumber  \label{lqr}\ell_{k,n} &= \sum_{j=0}^{N_c-1}\left\|{\tilde{\mathbf{y}}}_{k,n+j|n}-{\tilde{\mathbf{y}}}_{k,n+j}^{\mathrm{ref}} \right\|_\mathbf{Q}^2 + \left\|\mathbf{\tilde{u}}_{k,n+j|n}\right\|_\mathbf{R}^2\\&+\left\|{\tilde{\mathbf{y}}}_{k,n+N_c|n} -{\tilde{\mathbf{y}}}_{k,n+N_c}^{\mathrm{ref}}\right\|_{\mathbf{Q}_f}^2,  
\end{align}
where $N_c$, ${\tilde{\mathbf{y}}}_{k,n+j|n}$, and ${\tilde{\mathbf{y}}}_{k,n+j}^\mathrm{ref}$ denote the finite predictive control horizon, the predicted output at time slot $n+j$, and the known desired control ouput.  Here, the terms $\mathbf{Q}$, $\mathbf{R}$, and $\mathbf{Q}_f$ are the weighted matrices that allow for the prioritization of minimizing state deviations and control effort, respectively. To facilitate the considered system, an intuitive goal is to minimize the MPC cost \eqref{lqr}.

Meanwhile, given the limited on-board energy of UAS, it is required to harvest more power from the BS to prolong the lifetime.  Denote $P_{\max}$ as the maximum available power budget at the BS for each time slot. At the receiver side, it is evident that splitting more power to decode information can better guarantee the stabilizability, which in turn inevitably sacrifices the power that can be harvested. Hence, to achieve a balance between the control performance and the sustainability, we define a combined objective function given by $\frac{\sum_{k}^K\ell_{k,n}}{\sum_{k=1}^{K} P_{k,n}^{e}}
$, which measures the aggregate tracking loss per unit of harvested energy.
In our considered control-oriented SWIPT-enabled LAWN system, this objective is meaningful only if the downlink can sustain the information flow required for closed-loop stability. 
To this end, we have the following theorem:

\begin{theorem}
\label{lemma1}
Consider a linear and discrete system described by \eqref{spacestate}, a necessary condition for the system stabilizability with a feedback data rate of \( R_{k,n} \) is given by:
\begin{equation}
    R_{k,n} > \sum_{i=1} \max\left(0, \, \log_2 |\lambda_i'(\mathbf{\tilde{A}}_k)| \right)\triangleq R^{\mathrm{st}}_k, \forall n,
\end{equation}
where $\lambda_i'(\mathbf{\tilde{A}}_k)$ denotes the \(i'\)-th eigenvalue of the matrix \(\mathbf{\tilde{A}}_k\).
\begin{proof}
	Please refer to \cite{1310461}, \cite[Proposition 3.2]{1310480}.
\end{proof}
\end{theorem} 
Moreover, to comply with the stringent control delay requirement, we adopt a FBL transmission scheme in contrast to the conventional Shannon capacity. By doing so, \eqref{eq9} can be slightly modified as 
\begin{align}
    R_{k,n} = \log_2 \!\left( 1 +\mathrm{SINR}_{k,n} 
\right)-\sqrt{\frac{\mathcal{V}_{k,n}}{l_{k,n}}}\mathcal{Q}^{-1}\left(\varepsilon\right), \label{fblca}
\end{align}
where $l_{k,n}$ represents the blocklength of the transmitted control signals to the UAS $k$ in time slot $n$, $\varepsilon$ is the required block error rate, and $\mathcal{V}_{k,n}$ is the channel dispersion given by
\begin{equation}
\mathcal{V}_{k,n}=1-\left(1+\mathrm{SINR}_{k,n}\right)^{-2}.
\end{equation}
\revise{Since the physical-layer transmission duration scales with the adopted blocklength under a fixed symbol duration, a smaller \(l_{k,n}\) corresponds to a lower transmission latency requirement. Accordingly, the stringent delay requirement considered in this paper is reflected through the short blocklength regime, rather than through a full end-to-end delay analysis.}

Based on the discussions above, in time slot $n$, the ratio of the total control cost to the total harvested energy minimization problem, concerning the transmit beamforming, the PS ratios, and the control inputs optimization, can be formulated as follows:
\begin{subequations}
\begin{align}
\label{obj1p}  &\min_{\mathbf{W}_{k,n}, \mathbf{V}_n,\alpha_{k,n}, \mathbf{\tilde{U}}_{k,n}} \  \frac{\sum_{k}^K\ell_{k,n}}{\sum_{k=1}^{K} P_{k,n}^{e}} \\
        \mathrm{s.t.} \ & {\tilde{\mathbf{x}}}_{k,n+1}=\mathbf{\tilde{A}}_k{\tilde{\mathbf{x}}}_{k,n}+\mathbf{\tilde{B}}_k {\tilde{\mathbf{u}}}_{k,n},\ \mathbf{\tilde{y}}_{k,n}=\mathbf{D}_{k}\mathbf{\tilde{x}}_{k,n}, \label{p1a}\\ 
        & {\tilde{\mathbf{x}}}_{k,n|n}={\tilde{\mathbf{x}}}_{k,n}, \label{p1b}\\
&\mathbf{\tilde{U}}_{\min}\le\mathbf{\tilde{U}}_{k,n}\le \mathbf{\tilde{U}}_{\max}, \forall k,n, \label{p1c}\\
        &\log_2 \!\left( 1 +\mathrm{SINR}_{k,n} 
\right)-\sqrt{\frac{\mathcal{V}_{k,n}}{l_{k,n}}}\mathcal{Q}^{-1}\left(\varepsilon\right)\ge R^{\mathrm{st}}_k,\forall k, n,   \label{p1d}\\
&\sum_{k=1}^K \operatorname{tr}\left(\mathbf{W}_{k,n}\right)+\operatorname{tr}(\mathbf{V}_n) \leq P_{\max },\forall n, \label{p1e}\\
&  P_{k,n}^{e}\ge  P_{\min}^{e}, \forall k,n, \label{p1f}\\
& \mathrm{Rank}(\mathbf{W}_{k,n})\le 1, \mathbf{W}_{k,n},\mathbf{V}_n\succeq\mathbf{0},\forall k,n,  \label{p1g}\\
& 0 \le \alpha_{k,n} \le 1 , \forall k,n,  \label{p1h}
\end{align} \label{p1}%
\end{subequations}
where $\mathbf{\tilde{U}}_{k,n|n}=[\mathbf{\tilde{u}}_{k,n}^\top, \cdots, \mathbf{\tilde{u}}_{k,n+N_c|n}^\top]^\top\in \mathbb{R}^{2N_c}$ denotes the stacked control input sequence over the entire prediction horizon, with $\mathbf{\tilde{U}}_{\min}$ and $\mathbf{\tilde{U}}_{\max}$ being the minimum and maximum value of the control input, respectively. According to the MPC principle,  although a control sequence is optimized over the entire prediction horizon, only the first component, i.e., $\mathbf{\tilde{u}}_{k,n}$ is applied to ensure that the system state closely approaches the desired reference value. 
Constraints \eqref{p1a} and \eqref{p1b} enforce the system dynamics based on the linearized state-space model, while 
\eqref{p1c} imposes actuator feasibility by bounding the stacked control sequence over the prediction window. \revise{Constraint  
\eqref{p1d} establishes the cross-layer link between communication and control. Specifically, the effective achievable rate under finite blocklength transmission must exceed the minimum stabilizing rate to ensure closed-loop stabilizability. Therefore, communication degradation may directly reduce the achievable rate and consequently limit control performance.}  \eqref{p1e} imposes the maximum available transmit power per time slot at the BS. 
In \eqref{p1f}, $P_{\min}^e$ represents the minimum required harvested power for each UAS, while \eqref{p1g} ensures that $\mathbf{W}_{k,n}$ and $\mathbf{w}_{k,n}\mathbf{w}_{k,n}^H$ are equivalent.
Constraint \eqref{p1h} enforces that the received PS ratio for each UAS is confined within the interval $[0,1]$, ensuring a valid allocation between energy harvesting and information decoding.

It is observed that \eqref{p1} is intractable to solve due to the following reasons. First, problem \eqref{p1} is non-convex with the non-convexity arising from the fractional structure of the objective function as well as constraints \eqref{p1d} and \eqref{p1g}, which leads to finding the global optimal solution of \eqref{p1} is challenging. On the other hand, problem \eqref{p1} involves a dynamic programming structure due to the sequential nature of the control input optimization. Specifically, the control inputs over the entire prediction horizon are interdependent, and the optimization must account for the system's evolution over time. This introduces a high level of complexity, as the solution requires solving an optimization problem for each predicted step while considering the entire horizon.  Consequently, directly solving \eqref{p1} with traditional optimization methods is computationally expensive and may lead to intractability. Moreover, in the scenario with multiple no-fly zones, the reference trajectory should be carefully devised to ensure safe and accurate flight, which further complicates the solution to \eqref{p1}. To overcome these challenges, we develop a computationally efficient method to address problem \eqref{p1} in what follows.

\section{Stream Function-Based Reference Trajectory Design }\label{SEC4}
In this section, we propose an ideal fluid flow-based collision-free reference trajectory generation corresponding to the original problem \eqref{p1}.
Consider that the entire horizontal airspace, denoted by $\mathcal{F}\in \mathbb{R}^2$, contains both $M$ closed no-fly zones and an open flight-permitted region. Then, the planned subset of the airspace can be mathematically represented as
\begin{align}
   \mathcal{U}=\left( \partial \mathcal{F} \bigcup \mathcal{F} \right)\backslash \mathcal{O}, 
\end{align}
where $\partial \mathcal{F} $ is the boundary of the plane and $\mathcal{O}=\{\mathcal{O}_{1}\cup \cdots\cup \mathcal{O}_{M}\}$ collects $M$ NFZs, denoted by $\mathcal{M}=\{1,\cdots,M\}$.
We now adopt an ideal fluid flow pattern to describe the flight behavior. Specifically, for every location $\mathbf{p'}=[x',y']^\top\in \mathcal{F}$, the ideal irrotational  fluid flow field is given by 
\begin{align}
    \mathcal{I}(\mathbf{z})=\phi(x',y')+\mathbf{i}\psi(x',y'),
\end{align}
where  $\mathbf{z} =x'+\mathbf{i}y'$ is a complex variable with $\mathbf{i}$ denoting the imaginary part. Here, $\phi(x', y')$ and $\psi(x', y')$ represent the potential function and stream function, respectively. It is highlighted that $\phi(x',y')$ and $\psi(x',y')$ are required to satisfy the following constraints, i.e.,
\begin{equation} \label{pde}
\begin{aligned}
\nabla^2\psi(x',y')&\triangleq\frac{\partial^2\psi(x',y')}{\partial x'^2}+\frac{\partial^2\psi(x',y')}{\partial y'^2}=0,\\\nabla^2\phi(x',y')& \triangleq\frac{\partial^2\phi(x',y')}{\partial x'^2}+\frac{\partial^2\phi(x',y')}{\partial y'^2}=0,\\\frac{\partial\phi (x',y')}{\partial x'}&=\frac{\partial\psi(x',y')}{\partial y'},\ \quad\frac{\partial\phi(x',y')}{\partial y'}=-\frac{\partial\psi(x',y')}{\partial x'},
\end{aligned}
\end{equation}
in which the first two equations are the Laplace partial differential equations (PDE), while the last one is the Cauchy-Riemann condition. According to \cite{rastgoftar2019physics}, the principle of fluid flow-based trajectory design lies in that the UASs are constrained to move along the stream curves, i.e.,
\begin{align}
\psi(x_{k},y_{k})=\psi(x_{k}^\mathrm{in},y_{k}^\mathrm{in}), \forall k,n. 
\end{align}
Here, $x_{k}^\mathrm{in}$ and $ y_{k}^\mathrm{in}$ represent the locations of the boundary point where UAS $k$ enters $\mathcal{F}$. 
The stream function has a gradient perpendicular to the velocity field, specifying the velocity components, i.e., \begin{align} \label{part}
    v_x'=\frac{\partial \psi}{\partial y'},\ v_y'=-\frac{\partial \psi}{\partial x'},
\end{align} along the $x$ and $y$-axis.
Assuming that the flow is spatially varying and time-invariant at any point outside the  
boundary $\partial\mathcal{F}$ and the geometry of $\mathcal{O}$ keeps constant over time. Then, the goal is converted to solve for the stream function and the potential function 
that satisfy \eqref{pde}. The flow functions are used to model the UAS traffic coordination, ensuring smooth, collision-free trajectories for the UAS. In this context, the sink and source flows are widely employed, whose complex potential can be written as \cite{zhou2022guidance,smith2025stream}
\begin{equation}\label{comple}
\mathcal{I}(\mathbf{z})=\frac{C_0}{2} \ln \left(x'^2+y'^2\right)+\mathbf{i} C_0 \operatorname{atan}\left(\frac{y'}{x'}\right),
\end{equation}
where the constant $|C_0|$ represents the strength of the flow. Note that $C_0>0$ or $C_0 < 0$ classify a source or sink, respectively. However, the $M$ NFZs should be effectively enclosed within the complex ideal flow field. To this end, we have the following circle theorem to represent a circular obstacle in the flow.
\begin{theorem}
 (Circle Theorem). For an irrotational flow of incompressible inviscid fluid in $\mathcal{F}$. The singularities of complex potential $\mathcal{I}(\mathbf{z})$  are all at a distance greater than $R$ from the point $\mathbf{b}$. If a circular cylinder, typified by its cross-section the circle $|\mathbf{z}-\mathbf{b}|=R$, is introduced into the flow, the new complex potential becomes
\begin{align}
    \mathcal{I}'(\mathbf{z}) = \mathcal{I}(\mathbf{z})+\mathcal{{I}^\dagger}\left( \mathbf{b}^\dagger + \frac{R^2}{\mathbf{z} - \mathbf{b}} \right). \label{coms}
\end{align} \label{thre2}
\begin{proof}
    Please see \cite[Theorem 2.4]{1241966}
\end{proof}
\end{theorem}
By leveraging the theorem, one can arrive at the stream function with a single no-fly zone. In particular, for the $m$-th circular $\mathcal{O}_m$ centered at $\mathbf{p}'_m=[x'_m,y_m']^\top$ with radius $R_m$, the independent stream function in a sink flow is expressed as
\begin{equation}
\begin{aligned}
    &\psi_m(x',y')\\&=\underbrace{-C_0\mathrm{atan}\left(\frac{y'}{x'}\right)}_{\psi_1(0,0)}+\underbrace{C_0\mathrm{atan}\left(\frac{\frac{R_m^2(y'-y'_m)}{\|\mathbf{p'}-\mathbf{p}'_m \|^2}+y'_m}{\frac{R_m^2(x'-x'_m)}{\|\mathbf{p'}-\mathbf{p}'_m \|^2}+x'_m}\right)}_{\psi(x_m',y_m')}.
\end{aligned}\label{poten}
\end{equation}
  In fact,  Eq. \eqref{poten}, generated by Theorem \ref{thre2}, is called a doublet and is essentially the limit as a singular sink/source. Eq. \eqref{poten} is often adopted for path planning when the destination is modeled as a sink flow.  In particular, $\psi_1(0,0)$ is the stream induced by the destination located at the origin and $\psi(x'_m,y'_m)$ denotes the stream function components for the obstacle. \revise{ It is noted that although only circular obstacles are considered, the proposed navigation framework is readily extensible to arbitrary geometries, where irregularly shaped obstacles can be conservatively modeled using bounding circles to ensure a safety margin.}
\begin{figure}[t]
    \centering
\includegraphics[width=0.6\linewidth]{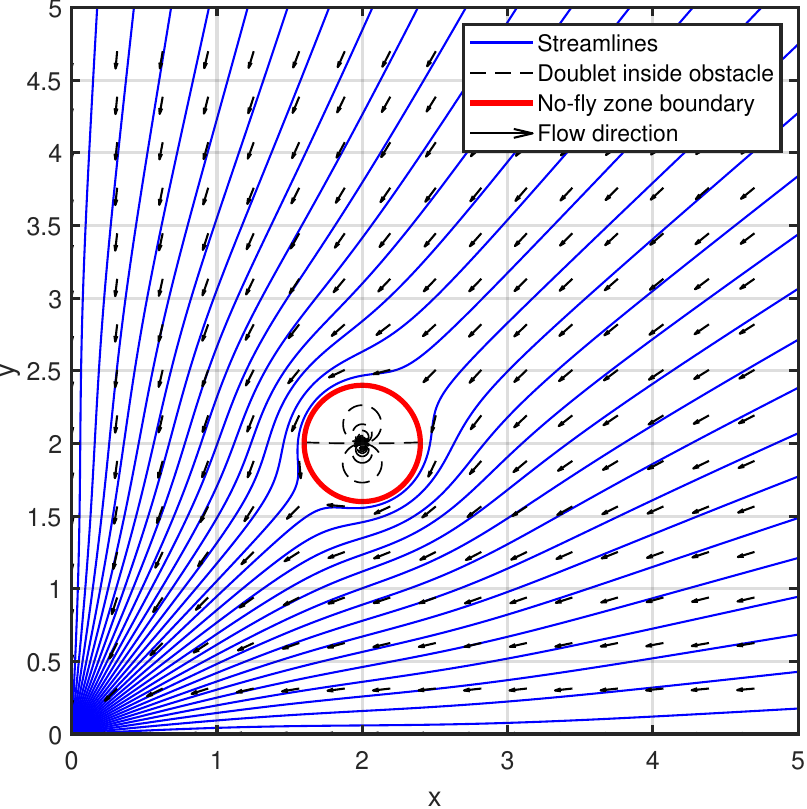}
    \caption{Static NFZ avoidance  via stream functions \eqref{poten}, where $R_m$=0.4 and the destination is located at (0,0). The UAS can fly along with the streamline to safely reach the destination.}
    \label{figpot}
\end{figure}
Fig. \ref{figpot} plots the no-fly zone $\mathcal{O}_m$ avoidance by leveraging the stream function \eqref{poten} with $R_m=0.4$. \revise{ For a UAS entering $\mathcal{F}$ through a boundary point, a safe reference path can be generated by following the streamline of the proposed guidance field toward the destination while avoiding contact with $\mathcal{O}_m$. It should be noted that \eqref{poten} is presented for the special case where the terminal point is located at the origin. More generally, for an arbitrary terminal point $\mathbf{p}'_{f}=[x'_{f},y'_{f}]^\top$, $\psi_1(0,0)$ can be replaced by
$
\psi_1(x'_f,y'_f)=-C_0\arctan\!\left(\frac{y'-y'_f}{x'-x'_f}\right).
$
With this modification, the proposed stream-function-based construction remains applicable to path generation from any prescribed start point to any assigned terminal point.}

In some cases, the NFZs may be mobile, e.g., temporarily restricted regions generated by surveillance aircraft or emergency response vehicles that dynamically change their positions over time. To avoid the collision with multiple mobile zones, it is required to add an additional vortex
flow centering at $\mathbf{p}_m$, which can be expressed as 
 \begin{equation} \label{vertex}
\psi_{\mathbf{v}_m}=C_{\mathbf{v}_m}S(\Theta_{m})\left|\mathbf{v}_m\right|\ln(\left|\mathbf{p}'-\mathbf{p}_m\right|^2),
 \end{equation}
 where $\mathbf{v}_m$ denotes the velocity of $\mathcal{O}_m$, $C_{\mathbf{v}_m}$ is a positive constant, and $S(\Theta_{m})$ is an indicating function given by 
 \begin{equation}
 S(\Theta_{m})=\left\{\begin{array}{cc}-1,&\mathrm{if~}\Theta_{1}<\Theta_{m}<\Theta_{2},\\1,&
\text{otherwise},\end{array}\right.
 \end{equation}
 where $\Theta_{m}\in [ -\pi,\pi)$ denotes the angle between the unit vectors $\frac{\mathbf{p'}_f-\mathbf{p}'}{|\mathbf{p'}_f-\mathbf{p}'|}$ and $\frac{\mathbf{v}_m}{|\mathbf v_m|}$. The terms $\Theta_1$ and $\Theta_2$ are the tuning  parameters, both satisfying $\Theta_1,\Theta_2\in (0,\pi)$. Then, the stream function is given by 
 \begin{align}
 \psi_{m}(x',y')=\psi_1(x'_f,y'_f)+\psi(x'_m,y'_m)+\psi_{\mathbf{v}_m}. \label{poteng1}
\end{align}
\begin{remark}
Here, $S(\Theta_{m})$ determines whether the mobile NFZ $m$ is bypassed via a clockwise or counterclockwise avoidance maneuver. For instance, when the direction toward the destination, $\frac{\mathbf{p}'_{f}-\mathbf{p}'}{\lVert \mathbf{p}'_{f}-\mathbf{p}' \rVert}$, is aligned with the NFZ’s velocity direction, $\frac{\mathbf{v}_{m}}{\lVert \mathbf{v}_{m} \rVert}$, i.e., a head-on or overtaking encounter, a counterclockwise vortex component is introduced, guiding the UAS to deviate to the right of the moving NFZ and ensuring collision-free passage.

\end{remark}
 Moreover, if a UAS is assigned to travel from a designated start location $\mathbf{p}'_{0}=[x'_{0},y'_{0}]^\top$ to $\mathbf{p}'_{f}$, one can further utilize the source flow to symbolize the start location  $\mathbf{p}'_{0}$ based on \eqref{comple}, leading to \eqref{poteng1} extending with an additional term $-\psi_1(x'_0,y'_0)$.
 \revise{In practical deployments, the NFZ region used in the stream-function construction can incorporate a safety margin by enlarging the prohibited area according to the required minimum separation distance. In this way, the generated guidance field ensures that the UAV trajectory maintains a safe distance from the actual NFZ boundary.}
 However, it is observed that the stream function is generated by a single NFZ. In scenarios where there are multiple circular regions, the Laplace’s equation with multiple boundary conditions must be solved to obtain the stream function, which is analytically intractable\cite{1241966}. 
 As a compromise, an addition method is applied \cite{sullivan2003using}, where the key lies in that the influence of each NFZ is approximately independent as long as they
are not nearly touching one another. Consequently, the stream function of multiple NFZs in the airspace is given by
\begin{align}
  \psi(x',y')= \sum_{m=1}^{M} \psi_m(x',y'). \label{totpon1}
\end{align}
By applying \eqref{totpon1} to generate the trajectory for each UAS, it is readily to obtain the reference state output $\mathbf{\tilde{y}}^\mathrm{ref}_{k,n}$ associated with UAS $k$ relying on \eqref{part}. To sum up, the streamline generation is outlined in \textbf{Algorithm \ref{alg1}}. \revise{Specifically, the proposed stream function is composed of three components, namely, a destination-oriented term that drives the UAS toward the assigned terminal point, an obstacle-induced term that enforces the non-penetrability of the NFZ boundary, and a vortex term introduced to capture the mobility of the NFZ.}
\begin{algorithm}[t]
  \caption{Stream Function Generation Algorithm}\label{alg1}
  \begin{algorithmic}[1] 
    \State \textbf{Input: $\mathbf{p}'_m, \mathbf{p'}, \mathbf{q}_k,\mathbf{p}'_f,\psi^0=0$} 
    \State \textbf{Parameter:} $R_m,C_0,M,K$
    \State \textbf{Output:}  $\psi(x',y')$
     \For{ $k\in \mathcal{K}$} 
      \While{ $\mathbf{q}_k\in \mathcal{F}$} 
      \State $\mathbf{p'} \gets \mathbf{q}_k$
       \For{ $m\in \mathcal{M}$} 
      \State $\psi(x',y')\gets \psi_m(x',y')+\psi^0$
      \State $\psi^0\gets \psi(x',y')$
      \EndFor
     \EndWhile
    \EndFor
  \end{algorithmic}
\end{algorithm}

\section{Efficient Algorithm for Solving \eqref{p1}} \label{SEC5}
With the reference control output $\mathbf{\tilde{y}}^\mathrm{ref}_{k,n}$ at hand, we now present an iterative algorithm to obtain a suboptimal solution to problem \eqref{p1}. The coupling between the transmit beamforming $\mathbf{W}_{k,n}, \mathbf{V}_{n}$ and the PS ratio $\alpha_{k,n}$, together with the fractional structure of the objective function, makes direct optimization challenging. Although there are some effective methods to handle the fractions \eqref{obj1p}, e.g., the Dinkbach approach \cite{shen2018fractional,9199556},  it is observed that the control input variable $\mathbf{\tilde{U}}_{k,n}$ is independent of $\{\mathbf{W}_{k,n},\mathbf{V}_n, \alpha_{k,n}\}$, which enables a separable optimization structure.
 Then, a two-step optimization framework can be proposed, in which the control input $\mathbf{\tilde{U}}_{k,n}$ is first optimized based on the reference $\mathbf{\tilde{y}}^\mathrm{ref}_{k,n}$. Subsequently, an AO algorithm is employed to jointly optimize the transmit beamforming $\mathbf{W}_{k,n}, \mathbf{V}_{n}$ and the PS ratio $\alpha_{k,n}$ in an iterative manner.

\subsection{Step 1: Control Input Design}

\revise{Given the reference control output \(\mathbf{\tilde{y}}_{k,n}^{\mathrm{ref}}\), the control-design subproblem of \eqref{p1} with respect to \(\{\mathbf{\tilde{U}}_{k,n}\}\) can be written as
\begin{align} \label{p2}
    \min_{\mathbf{\tilde{U}}_{k,n}} \quad \sum_{k=1}^{K} \ell_{k,n}
    \qquad
    \mathrm{s.t.}\quad \eqref{p1a}\text{--}\eqref{p1c}.
\end{align}
It is worth noting that problem \eqref{p2} is formulated directly in terms of the control input \(\mathbf{\tilde{u}}_{k,n}\). As such, it does not explicitly penalize or constrain the variation of the control input between two consecutive sampling instants. Consequently, the resulting control sequence may exhibit abrupt changes, which can impair control smoothness and reduce practical implementability. To address this issue, we introduce an augmented state by incorporating the previously applied control input into the system state, namely,
\begin{equation}
    \mathbf{\hat{x}}_{k,n}
=
\big[\mathbf{\tilde{x}}_{k,n}^{\top},\,
\mathbf{\tilde{u}}_{k,n-1}^{\top}\big]^{\top}.
\end{equation}
With this construction, the control design can be reformulated in terms of the control increment
$\Delta \mathbf{\tilde{u}}_{k,n}
\triangleq
\mathbf{\tilde{u}}_{k,n}-\mathbf{\tilde{u}}_{k,n-1},$
which enables a more explicit treatment of input variation and facilitates smoother control updates. The corresponding augmented state-space model is given by \eqref{newspace1} at the top of next page.
\begin{figure*}[ht]
\revise{\begin{align} \nonumber \hat{\mathbf{x}}_{k,n+1}&=\begin{bmatrix} \tilde{\mathbf{A}}_k\tilde{\mathbf{x}}_{k,n}+\mathbf{\tilde{B}}_k\mathbf{\tilde{u}}_{k,n}\\ \mathbf{\tilde{u}}_{k,n} \end{bmatrix}=\begin{bmatrix} \tilde{\mathbf{A}}_k\tilde{\mathbf{x}}_{k,n}+\mathbf{\tilde{B}}_k\mathbf{\tilde{u}}_{k,n-1}+\mathbf{\tilde{B}}_k\mathbf{\tilde{u}}_{k,n}-\mathbf{\tilde{B}}_k\mathbf{\tilde{u}}_{k,n-1}\\ \mathbf{\tilde{u}}_{k,n-1}+\mathbf{\tilde{u}}_{k,n}-\mathbf{\tilde{u}}_{k,n-1} \end{bmatrix}\\&=\begin{bmatrix} \mathbf{\tilde{A}}_k &\mathbf{\tilde{B}}_k \\ \mathbf{0}&\mathbf{I} \end{bmatrix} \begin{bmatrix} \tilde{\mathbf{x}}_{k,n}\\\tilde{\mathbf{u}}_{k,n-1} \end{bmatrix}+\begin{bmatrix} \mathbf{\tilde{B}}_k\\ \mathbf{I} \end{bmatrix} \big(\tilde{\mathbf{u}}_{k,n}-\tilde{\mathbf{u}}_{k,n-1}\big)\triangleq\mathbf{\hat{A}}_k\mathbf{\hat{x}}_{k,n}+\mathbf{\hat{B}}_k \Delta \mathbf{\tilde{u}}_{k,n}.\label{newspace1} \end{align}} \hrule \end{figure*}
Accordingly, the output equation can be rewritten as
$\mathbf{\hat{y}}_{k,n}=\mathbf{\hat{D}}_k \mathbf{\hat{x}}_{k,n},$
where
$\mathbf{\hat{D}}_k
=
\begin{bmatrix}
\mathbf{D}_k & \mathbf{0}
\end{bmatrix}
\in \mathbb{R}^{4\times 6}.$
This augmented formulation allows the controller to optimize the control increment rather than the absolute control input, thereby improving the continuity of the control sequence and making the resulting control actions more suitable for practical implementation.}

Then, the predictive state variable over the entire horizon $N_c$ can be expressed as 
\begin{equation}
    \begin{aligned}
\hat{\mathbf{x}}_{k,n|n} &= \hat{\mathbf{x}}_{k,n} \\
\hat{\mathbf{x}}_{k,n+1|n}
&= \hat{\mathbf{A}}_{k}\hat{\mathbf{x}}_{k,n} +  \hat{\mathbf{B}}_{k}\Delta\mathbf{u}_{k,n|n} \\
\hat{\mathbf{x}}_{k,n+2|n}
&= \hat{\mathbf{A}}_{k}^{2}\hat{\mathbf{x}}_{k,n}
+ \hat{\mathbf{A}}_{k}\hat{\mathbf{B}}_{k}\Delta\mathbf{u}_{k,n|n}
+\hat{\mathbf{B}}_{k}\Delta\mathbf{u}_{k,n+1|n} \\
&\ \ \cdots \\
\hat{\mathbf{x}}_{k,n+ N_c|n}
&= \hat{\mathbf{A}}_{k}^{ N_c}\hat{\mathbf{x}}_{k,n}
+ \hat{\mathbf{A}}_{k}^{ N_c-1}\hat{\mathbf{B}}_{k}\Delta\mathbf{u}_{k,n|n} + \cdots\\
&+\hat{\mathbf{B}}_{k}\Delta\mathbf{u}_{k,n+ N_c-1|n}
\end{aligned}
\end{equation}

Denoting the stacked predictive control input and output vector as $\Delta\mathbf{\tilde{U}}_{k,n}=[\Delta \mathbf{\tilde{u}}_{k,n|n}^\top,\cdots,\tilde{\mathbf{u}}_{k,n+N_c-1|n}^\top]^\top \in\mathbb{R}^{2N_c}$ and  $\mathbf{\hat{Y}}_{k,n}=[\mathbf{\hat{y}}_{k,n|n}^\top,\cdots,\mathbf{\hat{y}}_{k,n+N_c|n}^\top]^\top  \in\mathbb{R}^{4N_c+4}$, respectively, we then have 
\begin{align}
\mathbf{\hat{Y}}_{k,n}=\mathbf{S}_k\mathbf{\hat{x}}_{k,n}+\mathbf{T}_k\Delta\mathbf{\tilde{U}}_{k,n}, \label{finalste}
\end{align}
in which $\mathbf{S}_k=[\mathbf{\hat{D}}_k,\mathbf{\hat{D}}_k\mathbf{\hat{A}}_k, \mathbf{\hat{D}}_k\mathbf{\hat{A}}_k^2, \cdots,\mathbf{\hat{D}}_k\mathbf{\hat{A}}_k^{N_c}]^\top\in{\mathbb{R}}^{(4N_c+4)\times6}$ and $\mathbf{T}_k$ follows that 
\begin{equation}
\mathbf{T}_k=\left[\begin{array}{ccc}
\mathbf{0} &\cdots&\mathbf{0}\\
\hat{\mathbf{D}}_k\hat{\mathbf{B}}_k & \cdots & \mathbf{0} \\
\hat{\mathbf{D}}_k\hat{\mathbf{A}}_k \hat{\mathbf{B}}_k & \cdots & \mathbf{0} \\
\vdots & \ddots & \vdots \\
\hat{\mathbf{D}}_k\hat{\mathbf{A}}_k^{N_p-1} \hat{\mathbf{B}}_k & \cdots & \hat{\mathbf{D}}_k\hat{\mathbf{B}}_k
\end{array}\right]\in\mathbb{R}^{(4N_c+4)\times2N_c}.
\end{equation}
Now, the cost function \eqref{lqr} can be rewritten as
\begin{align}
\ell_{k,n} = \left\|{\hat{\mathbf{Y}}}_{k,n}-{\hat{\mathbf{Y}}}_{k,n}^{\mathrm{ref}} \right\|_\mathbf{\hat{Q}}^2 + \left\|\mathbf{\Delta{\tilde{U}}}_{k,n}\right\|_\mathbf{\hat{R}}^2. \label{38}
\end{align}
Here, $\mathbf{\hat{Q}}=\mathrm{diag}(\mathbf{Q},\cdots,\mathbf{Q},\mathbf{Q}_f)$ and $\mathbf{\hat{R}}=\mathbf{R}\otimes\mathbf{I}_{N_c} $ represent the reconstructed weight matrices. The term ${\hat{\mathbf{Y}}}_{k,n}^{\mathrm{ref}}\in\mathbb{R}^{4N_c+4} $ denotes the stacked reference control output over the entire predictive horizon. Plugging \eqref{finalste} into \eqref{38}, we have the following results, i.e.,
\begin{equation}
\begin{aligned}
\ell_{k,n}
&=\Delta\tilde{\mathbf{U}}_{k,n}^\top(\mathbf{T}_k^\top\hat{\mathbf{Q}}\mathbf{T}_k+\hat{\mathbf{R}})\Delta\tilde{\mathbf{U}}_{k,n}\\&+2\mathbf{e}_{k,n}^\top\hat{\mathbf{Q}}\mathbf{T}_k\Delta\tilde{\mathbf{U}}_{k,n}+\mathbf{e}_{k,n}^\top \hat{\mathbf{Q}} \mathbf{e}_{k,n},
\end{aligned} \label{finallqr}
\end{equation}
where $\mathbf{e}_{k,n}=\mathbf{S}_k\mathbf{\hat{x}}_{k,n}-\hat{\mathbf{Y}}_{k,n}^{\mathrm{ref}}$. Denoting $ \mathbf{H}_\mathrm{e} = 2\mathbf{T}_k^\top \hat{\mathbf{Q}} \mathbf{T}_k + \hat{\mathbf{R}}$ and $\mathbf{f}^\top = 2 \mathbf{e}_k^\top \hat{\mathbf{Q}} \mathbf{T}_k$, problem \eqref{p2} can be converted into a standard QP problem, which is given by

\begin{subequations}
    \begin{align}
    \min_{\mathbf{\Delta \tilde{U}}_{k,n}} &\frac{1}{2} \mathbf{\Delta \tilde{U}}_{k,n}^\top\mathbf{H}_\mathrm{e}\mathbf{\Delta \tilde{U}}_{k,n}+\mathbf{f}^\top \mathbf{\Delta \tilde{U}}_{k,n}\\
  &\mathrm{s.t.} \ \mathbf{\Delta \tilde{U}}_{k,n}^{\min}\le \mathbf{\Delta \tilde{U}}_{k,n}\le \mathbf{\Delta \tilde{U}}_{k,n}^{\max},
    \end{align} \label{p2-1}%
\end{subequations}
with $\mathbf{\Delta \tilde{U}}_{k,n}^{\max}$ and $\mathbf{\Delta \tilde{U}}_{k,n}^{\min}$ being the maximum and minimum values of the increment control quantities. As observed, problem \eqref{p2-1} is convex, and thereby, can be readily solved by standard tools, e.g., CVX\cite{grant2008cvx}.
\subsection{Step 2: AO Approach }
\subsubsection{Transmit Beamforming Design}\label{IVC}
Given the PS ratio $\{\alpha_{k,n}\}$, the subproblem of \eqref{p1} for optimizing the transmit beamforming $\{\mathbf{W}_{k,n}, \mathbf{V}_n \}$ can be expressed as
\begin{subequations}
\begin{align}
 \nonumber  &\max_{\mathbf{W}_{k,n}, \mathbf{V}_n } \  \sum_{k=1}^{K} P_{k,n}^{e} \\
        \mathrm{s.t.} \ & \eqref{p1d}-\eqref{p1g}.
\end{align} \label{p3}%
\end{subequations}
Problem \eqref{p3} is still a non-convex optimization problem and the non-convexity arises from  \eqref{p1d} and \eqref{p1g}. To circumvent this challenge, we introduce an auxiliary variable set $\{ \zeta_{k,n}, \forall k,n \}$, where $\zeta_{k,n}$ serves as a lower bound on the SINR for UAS $k$ in time slot $n$, i.e.,
\begin{align}
&\frac{\alpha_{k,n}  \mathrm{tr}\left(\mathbf{W}_{k,n} \mathbf{H}_{k,n}\right)} {\alpha_{k,n}\left(\mathrm{tr}\left(\sum_{k' \neq k} \mathbf{W}_{k',n} \mathbf{H}_{k,n} + \mathbf{V}_n\mathbf{H}_{k,n} \right)+ \sigma_k^2\right)+ \sigma_p^2} \geq \zeta_{k,n}. \label{constr1}
\end{align}
Therefore, constraint \eqref{p1d} can be transformed into
\begin{align}
\log_2 \left( 1 +\zeta_{k,n} \right)-\sqrt{\frac{1-\left(1+ \zeta_{k,n} \right)^{-2}}{l_{k,n}}}\mathcal{Q}^{-1}\left(\varepsilon\right)\ge R^{\mathrm{st}}_k, \label{constr2}
\end{align}
Clearly, the term $\sqrt{1-\left(1+ \zeta_{k,n} \right)^{-2}}$ is concave with respect to $\zeta_{k,n}$. Then, we employ the SCA technique to construct a tractable upper bound, i.e.,
\begin{equation}
\mathcal{V}_{k,n}^{\rm{ub}} \triangleq \sqrt{1-\left(1+ \zeta_{k,n}^{(l)} \right)^{-2}} + \frac{(\zeta_{k,n} - \zeta_{k,n}^{(l)})}{(1 + \zeta_{k,n}^{(l)})^3 \sqrt{1 - \frac{1}{(1+\zeta_{k,n}^{(l)})^2}}}. \label{constr3}
\end{equation}
By replacing the term $\sqrt{1-\left(1+ \zeta_{k,n} \right)^{-2}}$ in constraint \eqref{constr2} with its upper bound $\mathcal{V}_{k,n}^{\rm{ub}}$, 
 constraint \eqref{constr2} can be transformed into
\begin{align}
\log_2 \left( 1 +\zeta_{k,n} \right) - \mathcal{V}_{k,n}^{\rm{ub}} \frac{\mathcal{Q}^{-1}\left(\varepsilon\right)}{\sqrt{l_{k,n}}} \ge R^{\mathrm{st}}_k. \label{constr4}
\end{align}

Moreover, the left-hand side (LHS) of constraint \eqref{constr1} exhibits a fractional structure, with both the numerator and denominator involving optimization variables $\mathbf{W}_{k,n}, \mathbf{V}_n$, rendering constraint \eqref{constr1} non-convex. To this end, we again introduce an auxiliary variable set $\{ \delta_{k,n}, \forall k,n \}$, which serves as an upper bound for the denominator term, i.e.,
\begin{align}
&\alpha_{k,n}\left(\mathrm{tr}\left(\sum_{k' \neq k} \mathbf{W}_{k',n} \mathbf{H}_{k,n} + \mathbf{V}_n\mathbf{H}_{k,n} \right)+ \sigma_k^2\right)+ \sigma_p^2 \leq \delta_{k,n}. \label{constr5}
\end{align}
Therefore, constraint \eqref{constr1} can be rewritten as
\begin{align}
&\alpha_{k,n}  \mathrm{tr}\left(\mathbf{W}_{k,n} \mathbf{H}_{k,n}\right) \geq \zeta_{k,n} \delta_{k,n}. \label{constr6}
\end{align}
To address \eqref{constr6}, we have the following lemma:
\begin{lemma} \label{lemma2}
Given any two positive terms $x_{i}$ and $x_{j}$, the upper bound and lower bound of $x_{i}x_{j}$ can be expressed as
\begin{equation}
x_{i} x_{j} \leq \frac{1}{2} \left( \frac{x_{j}^{(l)}}{x_{i}^{(l)}} x_{i}^2 + \frac{x_{i}^{(l)}}{x_{j}^{(l)}} x_{j}^2 \right),
\end{equation}
\begin{equation}
x_{i} x_{j} \geq (x_{i}^{(l)} x_{j}^{(l)}) \left( 1 + \ln x_{i} + \ln x_{j} - \ln x_{i}^{(l)} - \ln x_{j}^{(l)} \right),
\end{equation}
where $x_{i}^{(l)}$ and $x_{j}^{(l)}$ are the given local points in the $l$-th iteration.
\begin{proof}
	Please refer to \cite{7547360}.
\end{proof}
\end{lemma}
According to Lemma \ref{lemma2}, we can obtain
\begin{align}
\zeta_{k,n} \delta_{k,n} \leq \frac{1}{2} \left( \frac{\zeta_{k,n}^{(l)}}{\delta_{k,n}^{(l)}} \delta_{k,n}^2 + \frac{\delta_{k,n}^{(l)}}{\zeta_{k,n}^{(l)}} \zeta_{k,n}^2 \right),
\end{align}
where $\delta_{k,n}^{(l)}$ is the local point in the $l$-th iteration. Therefore, constraint \eqref{constr5} can be transformed into 
\begin{align}
&\alpha_{k,n}  \mathrm{tr}\left(\mathbf{W}_{k,n} \mathbf{H}_{k,n}\right) \geq \frac{1}{2} \left( \frac{\zeta_{k,n}^{(l)}}{\delta_{k,n}^{(l)}} \delta_{k,n}^2 + \frac{\delta_{k,n}^{(l)}}{\zeta_{k,n}^{(l)}} \zeta_{k,n}^2 \right). \label{constr7}
\end{align}

Since the rank constraint \eqref{p1g} is inherently difficult to transform into a convex form, we resort to applying the SDR technique. Consequently, problem \eqref{p3} can be reformulated as
\begin{subequations}
\begin{align}
 \nonumber  \max_{\mathbf{W}_{k,n}, \mathbf{V}_n, \zeta_{k,n}, \delta_{k,n}} \  &\sum_{k=1}^{K} P_{k,n}^{e} \\
        \mathrm{s.t.} \ & \eqref{p1e}, \eqref{p1f}, \eqref{constr4}, \eqref{constr5}, \eqref{constr7}, \\
        & \mathbf{W}_{k,n},\mathbf{V}_n\succeq\mathbf{0},\forall k,n. \label{p4b}
\end{align} \label{p4}%
\end{subequations}
Clearly, problem \eqref{p4} is a standard semidefinite programming (SDP) problem, which can be efficiently solved using convex optimization toolboxes such as CVX \cite{grant2008cvx}. Regarding the relaxed rank constraint $\mathrm{Rank}(\mathbf{W}_{k,n})\le 1$, if the obtained optimal solution $\mathbf{W}_{k,n}^{*}$ does not satisfy the rank-one condition, Gaussian randomization technique can be employed to construct a rank-one approximate solution \cite{5447068}. Fortunately, by analyzing the Karush-Kuhn-Tucker (KKT) optimality conditions of the relaxed problem \eqref{p4}, it can be shown that there always exists a rank-one optimal solution $\mathbf{W}_{k,n}^{*}$. The complete proof is provided in Appendix \ref{proof1}.

\subsubsection{PS Ratio Design}
Given the transmit beamforming $\{\mathbf{W}_{k,n}, \mathbf{V}_n \}$, the subproblem of \eqref{p1} for optimizing the PS ratio $\{\alpha_{k,n}\}$ can be expressed as
\begin{equation}
\begin{aligned}
   &\max_{\alpha_{k,n}} \  \sum_{k=1}^{K} P_{k,n}^{e} \\
        \mathrm{s.t.} \ & \eqref{p1d}, \eqref{p1f}, \eqref{p1h}.
\end{aligned} \label{p5}%
\end{equation}
Due to constraint \eqref{p1d}, problem \eqref{p5} is still non-convex. Following the same approach as in Section \ref{IVC}, we can transform problem \eqref{p5} into the following problem \eqref{p6},
\begin{equation}
\begin{aligned}
  &\max_{\alpha_{k,n}, \zeta_{k,n}, \delta_{k,n}} \  \sum_{k=1}^{K} P_{k,n}^{e} \\
        \mathrm{s.t.} \ & \eqref{p1f}, \eqref{p1h}, \eqref{constr4}, \eqref{constr5}, \eqref{constr7}, 
\end{aligned} \label{p6}%
\end{equation}
Problem \eqref{p6} is now a standard convex optimization problem, which can be efficiently solved via CVX \cite{grant2008cvx}.

\revise{Based on the above analysis, problem \eqref{p1} admits an iterative solution framework. Specifically, Algorithm~\ref{alg1} is first adopted to generate reference trajectories via the stream-function-based method, thereby enabling the drones to approach their destinations safely without colliding with the NFZs. Building upon these reference trajectories, a two-step optimization scheme is then developed to decouple the design of control actions and wireless resource allocation.} In particular, we first address the control input over the entire predictive horizon $N_c$ by solving problem \eqref{p2-1}. Then, the beamforming vector is designed by addressing the SDR problem \eqref{p4} given $\{\mathbf{\tilde{U}
}_{k,n},\alpha_{k,n}\}$, Subsequently, we deal with problem \eqref{p6} to determine the optimal PS ratio with other variables remaining constant. To sum up, the overall algorithm for problem \eqref{p1} is summarized in {Algorithm \ref{alg2}}. 
\revise{We now discuss the computational complexity of Algorithms~\ref{alg1} and~\ref{alg2}. In particular,
Algorithm~\ref{alg1} generates the reference trajectories by evaluating the stream function of each NFZ along the sampled path points. Hence, its complexity is linear given by $\mathcal{O}(K M N_{\mathrm{ref}}).$ For Algorithm~\ref{alg2}, the control update in step~1 is a convex quadratic program with complexity
$\mathcal{O}\!\left(K(2N_c)^{3.5}\right).$
In step~2, the main complexity arises from the SDR-based beamforming subproblem, while the power-splitting update has much lower complexity. Let
$n_{\mathrm{s}} \triangleq (K+1)N_t^2+2K.$
Then, the overall complexity of Algorithm~\ref{alg2} can be summarized as
$\mathcal{O}\!\left(
K(2N_c)^{3.5}
+
I_{\mathrm{AO}}
\left(
n_{\mathrm{s}}^{3.5}
+
(3K)^{3.5}
\right)
\right),$
where \(I_{\mathrm{AO}}\) denotes the number of alternating iterations. }

\revise{We proceed to analyze the convergence behavior. Since problem \eqref{p2-1} is convex, the control input in step~1 is obtained optimally. Thus, the main convergence issue lies in the AO procedure in step~2. Define the objective value at the \(l\)-th iteration as $\Phi^{(l)}$. For fixed \(\{\alpha_{k,n}^{(l-1)}\}\), problem \eqref{p4} is a convexified beamforming subproblem, and the local tightness of the adopted SCA surrogates ensures
\begin{equation}
\Phi\!\left(\{\mathbf{W}_{k,n}^{(l-1)},\mathbf{V}_{n}^{(l-1)},\alpha_{k,n}^{(l-1)}\}\right)
\le
\Phi\!\left(\{\mathbf{W}_{k,n}^{(l)},\mathbf{V}_{n}^{(l)},\alpha_{k,n}^{(l-1)}\}\right).
\end{equation}
Similarly, for fixed \(\{\mathbf{W}_{k,n}^{(l)},\mathbf{V}_{n}^{(l)}\}\), solving problem \eqref{p6} yields
\begin{equation}
\Phi\!\left(\{\mathbf{W}_{k,n}^{(l)},\mathbf{V}_{n}^{(l)},\alpha_{k,n}^{(l-1)}\}\right)
\le
\Phi\!\left(\{\mathbf{W}_{k,n}^{(l)},\mathbf{V}_{n}^{(l)},\alpha_{k,n}^{(l)}\}\right).
\end{equation}
Hence, \(\Phi^{(l-1)}\le \Phi^{(l)}\), namely, the objective sequence is monotonically non-decreasing. Moreover, the objective is upper-bounded due to the limited available transmit power budget. Therefore, the sequence generated by Algorithm~\ref{alg2} converges to a finite limit, guaranteeing the convergence of original problem \eqref{p1}.  }
\begin{algorithm}[t]
  \caption{Overall Algorithm for Solving Problem \eqref{p1}}\label{alg2}
  \begin{algorithmic}[1] 
    \State \textbf{Input: ${P}_{\max}, {R}_{k}^{\mathrm{st}}, P^e_{\min},K,N_c,\mathbf{\Delta \tilde{U}}_{k,n}^{\max}$}, and $\mathbf{\Delta \tilde{U}}_{k,n}^{\min}$
    \State \textbf{Output:}  Optimal $\{\mathbf{W}^*_{k,n},\mathbf{V}^*_n, \alpha^*_{k,n},\mathbf{\tilde{U}}^*_{k,n}\}$
     \State Generate the reference trajectory via \textbf{Algorithm} \ref{alg1}.
    \State Obtain $\mathbf{\tilde{U}}_{k,n}$ by solving problem \eqref{p2-1}.
     \Repeat
     \State Obtain $\{\mathbf{{W}}_{k,n}, \mathbf{V}_n\}$ with other variables fixed by solving problem \eqref{p4}.
       \State Obtain $ \alpha_{k,n}$ with other variables fixed by solving problem \eqref{p6}.
     \Until{Converged}
     \State $\{\mathbf{W}^*_{k,n},\mathbf{V}^*_n, \alpha^*_{k,n},\mathbf{\tilde{U}}^*_{k,n}\} \gets \{\mathbf{W}_{k,n},\mathbf{V}_n, \alpha_{k,n},\mathbf{\tilde{U}}_{k,n}\}$
  \end{algorithmic}
\end{algorithm}

\section{Simulation Results}\label{SEC6}

In this section, we perform extensive simulations to validate the effectiveness of the proposed design. Unless otherwise specified, the considered scenario consists of a square low-altitude airspace $\mathcal{F}$ with dimensions $0.5 \, \text{km} \times 0.5 \, \text{km}$, where $K = 3$ UASs enter $\mathcal{F}$ from random points on the boundary $\partial\mathcal{F}$. All UASs share a common destination $\mathbf{p}_f = [1, 1]^\top$ m and operate at a flight height of $H_k = 20 \, \text{m}$. The BS is located at a fixed point $\mathbf{g} = [200, 200]^\top \, \text{m}$ at an altitude of $H_g = 5 \, \text{m}$, with $N_t = 6$ antennas spaced at $d = \lambda/2$. Once the UAS enters $\mathcal{F}$, the BS assigns a predefined reference trajectory based on the stream function and transmits the optimal control inputs to guide the UASs along the reference path. The receiver noise power is set to $\sigma_k^2 = \sigma_p^2 = -80 \, \text{dBm}$, and the energy conversion efficiency for each UAS is set to $\eta_k = 0.8$, $\forall k$. The channel gain at a unit reference distance is $\beta_0=-30$ dB and the maximum transmit power is $P_{\max} = 40 \, \text{dBm}$. For stream function generation, the flow strength and vortex strength coefficients are set to $C_0 = 5$ and $C_{\mathbf{v}_m} = 1.2$, respectively, with tuning parameters $\Theta_1 = \pi/6$ and $\Theta_2 = 5\pi/6$ \cite{zhou2022guidance}. Regarding the control, the length of predictive horizon is set to $N_c=10$, with the weighting matrices $\mathbf{Q}$, $\mathbf{R}$, and $\mathbf{Q}_f$ being identity matrices\cite{jin2025predictive}. 

\begin{figure}[t]
    \centering
\includegraphics[width=0.8\linewidth]{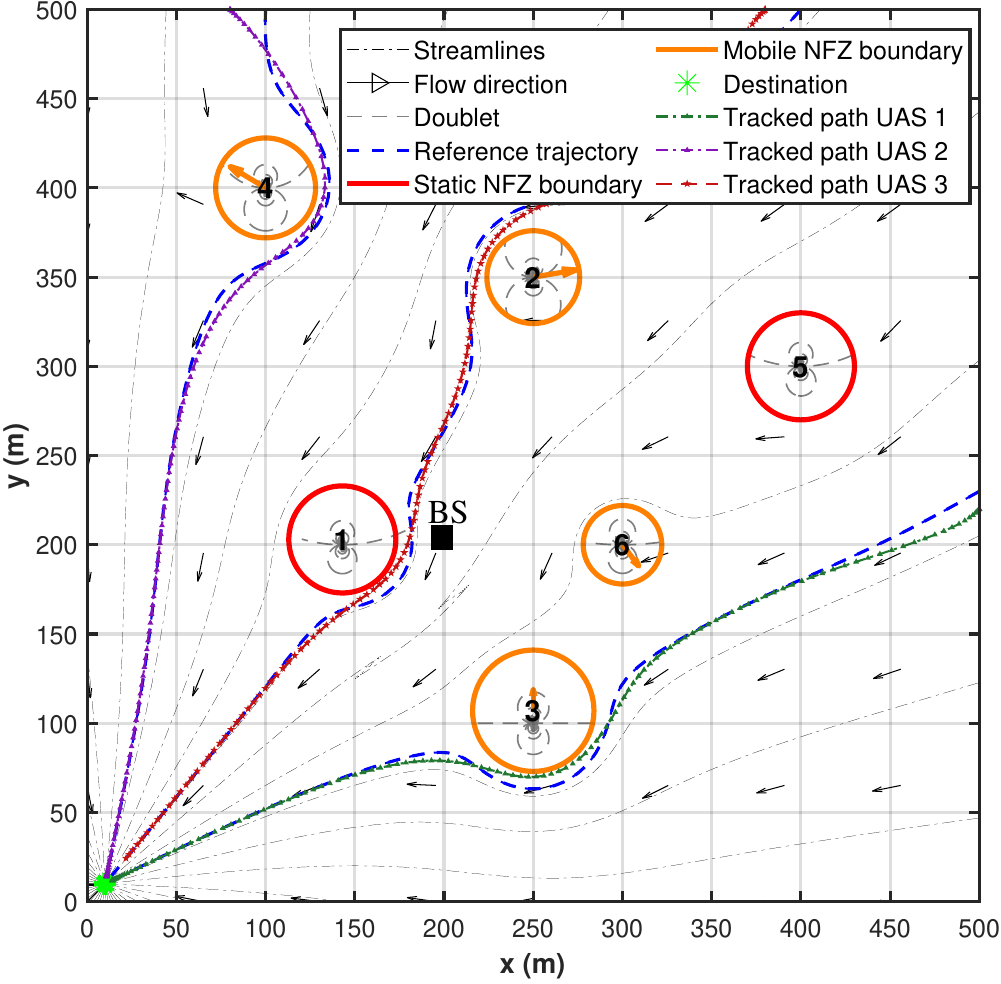}
    \caption{Reference trajectory generation and the corresponding tracking performance in the presence of both mobile and static NFZs with the BS at $\mathbf{g}=[200,200]^\top$ m.  The orange arrow indicates the NFZ's moving direction and velocity amplitude. }
    \label{fig3}
\end{figure}

\begin{table}[t]
\centering
\caption{NFZ Parameters}
\begin{tabular}{|c|c|c|c|c|}
\hline
\textbf{NFZs} & \textbf{Position}/m  & \textbf{Radius}/m & \textbf{Status} \\ \hline
\textbf{NFZ 1} &[140; 200] & 30 & Static \\ \hline
\textbf{NFZ 2} & [250; 350] & 26 & Mobile \\ \hline
\textbf{NFZ 3} & [250; 110] & 34 & Mobile \\ \hline
\textbf{NFZ 4} &[100; 400] & 28 & Mobile \\ \hline
\textbf{NFZ 5} & [400; 300] & 30 & Static \\ \hline
\textbf{NFZ 6} & [300; 200] & 22 & Mobile \\ \hline
\end{tabular} \label{TABLENFZ}
\end{table}
We begin by evaluating the NFZ avoidance performance of the proposed approach. Fig. \ref{fig3} demonstrates the reference trajectory generation and tracking performance of multiple UASs under the constraint of several NFZs. Specifically, we consider $M = 6$ NFZs, with their parameters provided in {TABLE \ref{TABLENFZ}}. As observed, the streamlines naturally avoid these NFZs due to the inherent properties of the stream function, which generates continuous, divergence-free paths that inherently bypass any restricted regions. Interestingly, in the case of {static NFZs}, UASs are able to maneuver around them from either side, as the flow remains predictable and steady. However, for {mobile NFZs}, the dynamic influence of the vortex induced by the moving NFZ compels the UAS to adjust its trajectory based on \eqref{vertex}. This leads to a curved path, as the flow becomes rotational, requiring the UAS to navigate around the NFZ to maintain safe flight. 

Subsequently, we evaluate the reference trajectory tracking performance. Once the UASs enter the region $\mathcal{F}$, available reference trajectories are assigned, beginning at positions $[500, 230]^\top$ m, $[100, 500]^\top$ m, and $[400, 500]^\top$ m in our considered scenarios. Although the UASs may not start precisely along the reference paths, the proposed MPC-based framework ensures gradual alignment with the desired trajectory over time. Notably, as the UASs approach mobile NFZs, tracking accuracy decreases, particularly when sharp turns are needed to avoid those dynamic NFZs. This phenomenon results from the sudden changes in flight dynamics caused by the mobile NFZs. Nevertheless, once the UASs pass the NFZ, the tracking error quickly diminishes.

\begin{figure}
    \centering
    \includegraphics[width=.9\linewidth]{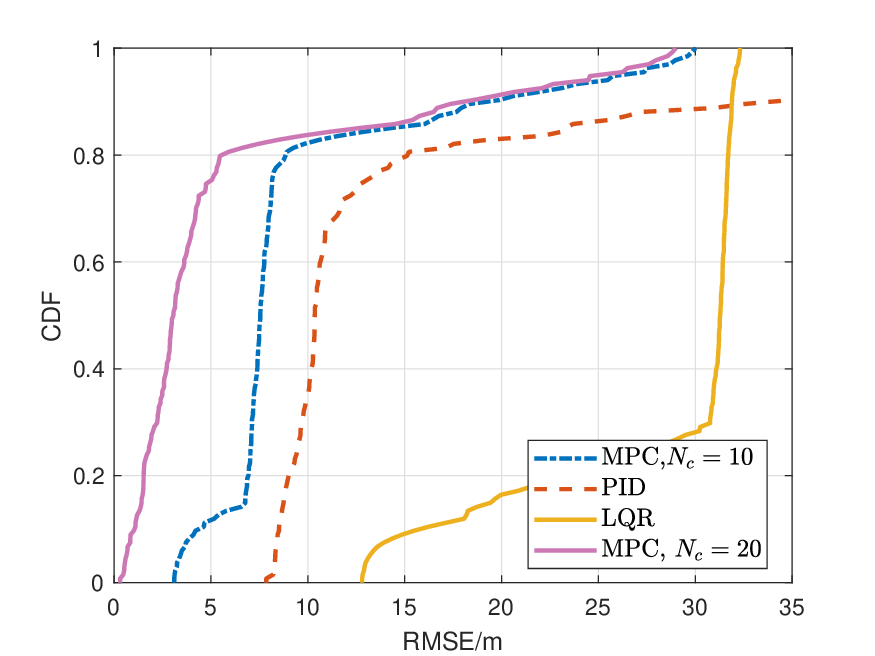}
    \caption{The CDF of trajectory tracking performance in terms of RMSE with various schemes.}
    \label{fig4}
\end{figure}

 To further emphasize the robustness and effectiveness of the proposed design, we compare our method with widely adopted control strategies, namely proportional-integral-derivative (PID) and linear-quadratic regulator (LQR)\cite{10702475}. As depicted in Fig. \ref{fig4}, the trajectory tracking performance of our MPC approach, as shown by the cumulative distribution function (CDF) of root mean square error (RMSE), significantly surpasses both PID and LQR. This superior performance can be attributed to the fact that both PID and LQR are offline methods that react solely to the current error, making them less effective at maintaining accurate tracking in dynamic environments. In contrast, MPC method optimizes control actions over a prediction horizon, enabling it to adapt to real-time changes, which makes it the most robust and precise method for controlling UASs in evolving airspace. Furthermore, the CDF of RMSE illustrates that with a longer prediction horizon of $N_c = 20$, the MPC approach consistently achieves the lowest RMSE values, as it anticipates future trajectory deviations and adjusts its control inputs accordingly.

\begin{figure}[t]
\centering
\includegraphics[width=0.9\linewidth]{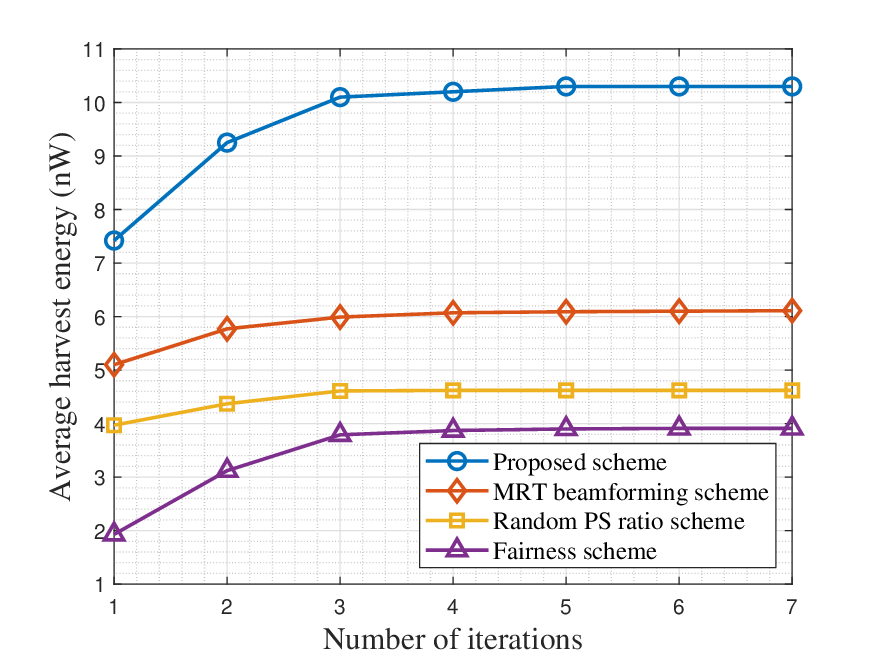}
\caption{The convergence behavior under different schemes.}
\label{convergence algorithm}
\end{figure}
Fig. \ref{convergence algorithm} illustrates the convergence behavior of our proposed design in terms of average harvested power across all time slots. In particular, we compare our method to various benchmark schemes, i.e.,
\begin{itemize}
    \item Fairness scheme: This scheme aims to optimize the objective function $\min \ \max \frac{\ell_{k,n}}{P_{k,n}^e}$ rather than \eqref{obj1p} to ensure UAS fairness.
    \item MRT beamforming scheme: This method adopts the maximum ratio transmission (MRT) technique, which maximizes the received signal power by aligning the transmission direction with the channel conditions. 
    \item Random PS ratio scheme: This scheme generates the PS ratio randomly without solving problem \eqref{p6}.
\end{itemize}
 It is observed that all the schemes can converge fast with no more than $10$ iterations. Moreover, we note that the random PS ratio scheme converges the fastest since it avoids solving all the subproblems related to PS ratio optimization.

\begin{figure}[t]
\centering
\includegraphics[width=0.9\linewidth]{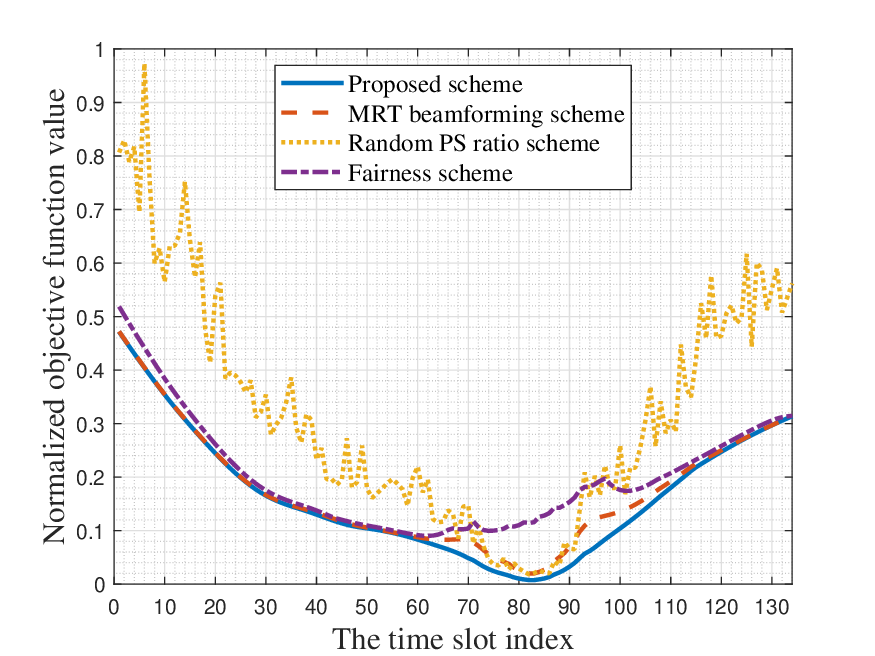}
\caption{The variation of the normalized objective function value.}
\label{objective value time}
\end{figure}

\begin{figure}[t]
\centering
\includegraphics[width=0.9\linewidth]{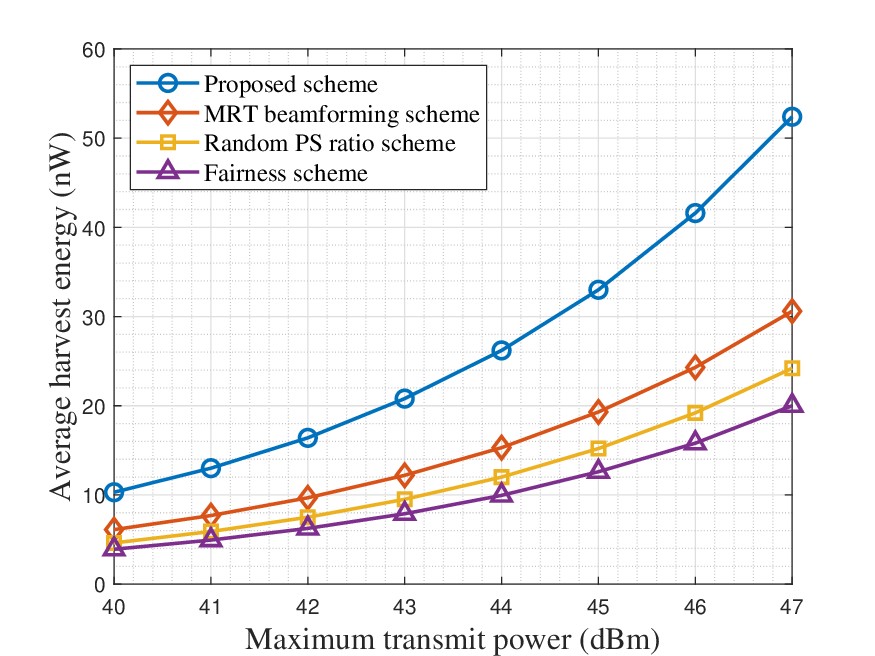}
\caption{The average harvested energy versus the maximum
 transmit power.}
\label{transmit power}
\end{figure}

\begin{figure}[t]
\centering
\includegraphics[width=0.9\linewidth]{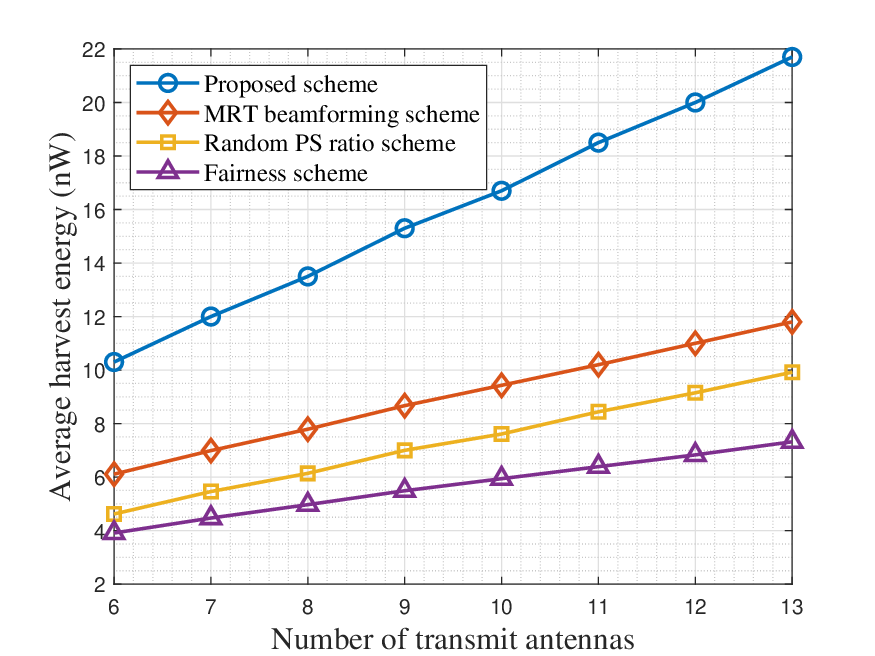}
\caption{The average harvested energy vs. transmit antenna numbers.}
\label{antenna number}
\end{figure}
\begin{figure}[t]
\centering
\includegraphics[width=0.9\linewidth]{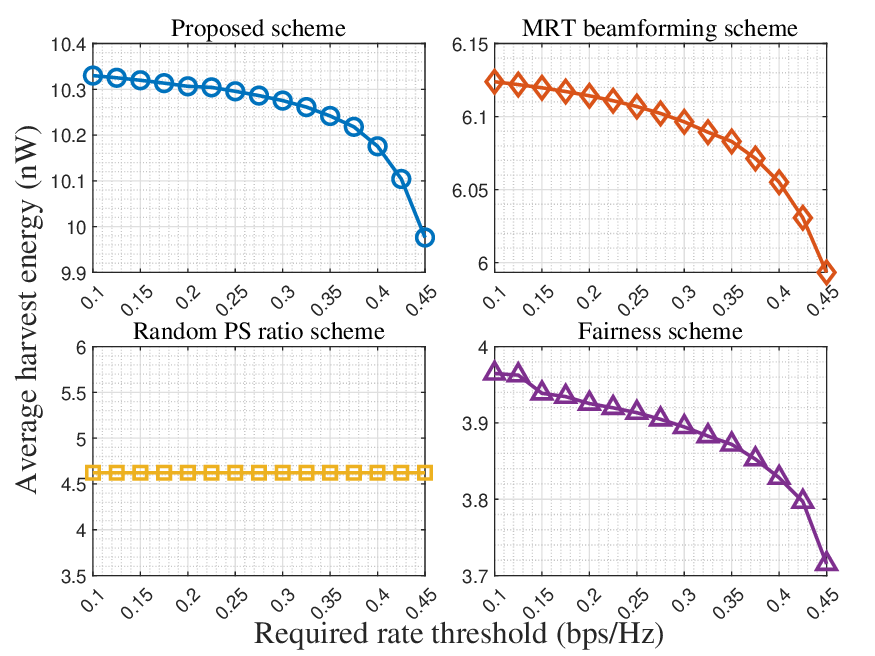}
\caption{The average harvested energy versus the required SE threshold.}
\label{rate threshold}
\end{figure}
In Fig. \ref{objective value time}, we illustrate the value variation of the normalized objective function \eqref{obj1p} over the entire duration, which captures the trade-off between control accuracy and energy sustainability in LAWN systems.  Initially, we see that the value for all methods decreases as the UASs approach the BS, where the communication quality and EH efficiency are optimized, enabling the UASs to closely follow the reference trajectory. However, as the UASs move further from the BS towards their destination, the degradation of both communication performance and energy efficiency leads to an increase in the objective function value, which exactly aligns with the trajectory depicted in Fig. \ref{fig3}. Notably, the random PS ratio scheme exhibits \textcolor{black}{the poorer performance} due to the absence of a properly designed power splitting mechanism, failing to balance control cost and energy sustainability. Additionally, the fairness scheme compromises by allocating more power to weaker UASs, leading to the performance degradation of the whole network.

 Fig. \ref{transmit power}  depicts the relationship between average harvested energy and transmit power $P_{\max}$. As expected, increasing the transmit power enhances energy harvesting for all methods, but the proposed scheme consistently outperforms the others due to the jointly optimized control and EH strategy. In contrast, the MRT beamforming scheme shows moderate improvement but falls short of the proposed approach due to its fixed beamforming technique. The fairness scheme and random PS ratio scheme yield lower EH performance. This phenomenon arises from the fact that the fairness scheme prioritizes equal distribution at the cost of each UAS, and the random PS ratio scheme lacks optimized power splitting. Similarly, Fig. \ref{antenna number} illustrates the relationship between average harvested energy and the number of transmit antennas for four different schemes. As the number of transmit antennas increases, all schemes show a corresponding increase in harvested energy, reflecting the benefits of additional antennas for improving communication efficiency and energy harvesting.
 
Fig. \ref{rate threshold} illustrates the non-trivial trade-off between the required SE threshold and the harvested energy. As the SE threshold becomes more stringent, the harvested energy decreases, revealing the inherent tension between ensuring control stability and maintaining EH efficiency. A higher SE threshold enhances communication reliability for closed-loop stability but simultaneously reduces the portion of received power available for EH, since more energy must be allocated to information decoding. It is also observed that the random PS ratio scheme is largely insensitive to variations in the SE threshold due to the absence of PS optimization, resulting in persistently inefficient EH performance. Furthermore, according to \eqref{fblca}, increasing the SE threshold necessitates a longer transmit blocklength. While a larger $R^\mathrm{st}_k$ improves the stabilizability of UAS $k$ as suggested by Theorem \ref{lemma1}, it also introduces additional risk of control delay, which may jeopardize timely trajectory tracking in practical implementations.

\section{Conclusion}\label{SEC7}
This paper investigated wireless control for SWIPT-enabled LAWNs within a dynamic airspace with mobile NFZs. The smooth and collision-free reference trajectories are first generated via stream function theory, followed by formulating a real-time minimization problem that balances reference trajectory tracking accuracy and energy sustainability by jointly optimizing transmit beamforming, PS ratios, and control inputs. The resulting non-convex problem was addressed through a two-stage approach, where the MPC framework and an iterative SDR-based algorithm are employed to solve predictive control inputs and EH strategy, respectively.   Simulation results demonstrated that the proposed design achieved superior tracking precision and harvested energy compared to benchmark schemes.

\begin{appendices}
\section{Proof of the Rank-One Property of $\{ \mathbf{W}_{k,n}^{*} \}$ } \label{proof1}
It can be readily verified that problem \eqref{p4} is convex and satisfies the Slater’s condition. Hence, there exists no duality gap between the primal and dual problems. By retaining only the terms associated with the transmit beamforming matrices $\mathbf{W}_{k,n}$, the corresponding Lagrangian of problem \eqref{p4} is given by
\begin{align}
& \mathbf{\mathcal{L}}\left( \mathbf{W}_{k,n}, \mu^{\rm{1}}_{n}, \mu^{\rm{2}}_{k,n}, \mu^{\rm{3}}_{k,n}, \mu^{\rm{4}}_{k,n}, \mathbf{\Lambda}_{k,n} \right) \nonumber \\
& = \sum_{k=1}^{K} P_{k,n}^{e} + \mu^{\rm{1}}_{n} \bigg( P_{\max} - \sum_{k=1}^K \operatorname{tr}\left(\mathbf{W}_{k,n}\right) -\operatorname{tr}(\mathbf{V}_n) \bigg)  \nonumber \\
& \quad + \sum_{k=1}^{K} \mu^{\rm{2}}_{k,n} \left(P_{k,n}^{e} - P_{\min}^{e} \right) + \sum_{k=1}^{K} \mu^{\rm{3}}_{k,n} \Bigg( \delta_{k,n} - \sigma_p^2 \nonumber \\
&\quad - \alpha_{k,n}\bigg(\mathrm{tr}\bigg(\sum_{k' \neq k} \mathbf{W}_{k',n} \mathbf{H}_{k,n} + \mathbf{V}_n\mathbf{H}_{k,n} \bigg)+ \sigma_k^2\bigg)  \Bigg) \nonumber \\
& \quad + \sum_{k=1}^{K} \mu^{\rm{4}}_{k,n} \Bigg( \alpha_{k,n}  \mathrm{tr}\left(\mathbf{W}_{k,n} \mathbf{H}_{k,n}\right) - \tfrac{1}{2} \tfrac{\zeta_{k,n}^{(l)}}{\delta_{k,n}^{(l)}} \delta_{k,n}^2 \nonumber \\
&\quad - \tfrac{1}{2}  \tfrac{\delta_{k,n}^{(l)}}{\zeta_{k,n}^{(l)}} \zeta_{k,n}^2  \Bigg) + \sum_{k=1}^{K} \mathrm{tr}\left(\mathbf{\Lambda}_{k,n}\mathbf{W}_{k,n}\right),
\end{align}
where $\{ \mathbf{\Lambda}_{k,n} \succeq \mathbf{0},\forall k \}$ denotes the dual matrix for the positive semi-definite constraint \eqref{p4b}, and $\{ \mu^{\rm{1}}_{n}, \mu^{\rm{2}}_{k,n}, \mu^{\rm{3}}_{k,n}, \mu^{\rm{4}}_{k,n}\}$ represents the dual variable set for constraints \eqref{p1e}, \eqref{p1f}, \eqref{constr5}, and \eqref{constr7}.
The KKT conditions are then given by
\begin{align}
\mathrm{tr}\left(\mathbf{\Lambda}_{k,n}\mathbf{W}_{k,n}\right) = \mathbf{0}, \label{proof1_1} 
\end{align}
\begin{align}
\dfrac{\partial \mathbf{\mathcal{L}}\left( \mathbf{W}_{k,n}, \mu^{\rm{1}}_{n}, \mu^{\rm{2}}_{k,n}, \mu^{\rm{3}}_{k,n}, \mu^{\rm{4}}_{k,n}, \mathbf{\Lambda}_{k,n} \right)}{\partial \mathbf{W}_{k,n}} = \mathbf{0}, \label{proof1_2} 
\end{align}
where 
\begin{align}
& \dfrac{\partial \mathbf{\mathcal{L}}\left( \mathbf{W}_{k,n}, \mu^{\rm{1}}_{n}, \mu^{\rm{2}}_{k,n}, \mu^{\rm{3}}_{k,n}, \mu^{\rm{4}}_{k,n}, \mathbf{\Lambda}_{k,n} \right)}{\partial \mathbf{W}_{k,n}} \nonumber \\
& = \sum_{k=1}^{K} \eta_k(1-\alpha_{k,n})(1+\mu^{\rm{2}}_{k,n}) \mathbf{H}_{k,n} - \mu^{\rm{1}}_{n} \mathbf{I}  \nonumber \\
& \quad + \sum_{k' \neq k} \mu^{\rm{3}}_{k',n} \alpha_{k',n} \mathbf{H}_{k',n} + \mu^{\rm{4}}_{k,n} \alpha_{k,n} \mathbf{H}_{k,n} + \mathbf{\Lambda}_{k,n} \nonumber \\
& = \tilde{\mathbf{H}}_{n} - \mu^{\rm{1}}_{n} \mathbf{I} + \mathbf{\Lambda}_{k,n},
\end{align}
with
\begin{align}
\tilde{\mathbf{H}}_{n} =& \sum_{k=1}^{K} \eta_k(1-\alpha_{k,n})(1+\mu^{\rm{2}}_{k,n}) \mathbf{H}_{k,n}  \nonumber \\
&+ \sum_{k' \neq k} \mu^{\rm{3}}_{k',n} \alpha_{k',n} \mathbf{H}_{k',n} + \mu^{\rm{4}}_{k,n} \alpha_{k,n} \mathbf{H}_{k,n}.
\end{align}

From \eqref{proof1_2}, we derive the expression $\mathbf{\Lambda}_{k,n} = \mu^{\rm{1}}_{n} \mathbf{I} - \tilde{\mathbf{H}}_{n}$.
Recall that $\mathbf{H}_{k,n}= \mathbf{h}_{k, n} \mathbf{h}_{k,n}^H$, which is an outer product of the vector $\mathbf{h}_{k, n}$ with itself. As such, $\mathbf{H}_{k,n}$ is a rank-one Hermitian matrix, i.e., $\mathrm{Rank}(\mathbf{H}_{k,n}) = 1$. Since $\tilde{\mathbf{H}}_{n}$ is constructed as a linear combination of such rank-one matrices, it is also a rank-one matrix.
By the rank inequality for matrices, we have
\begin{align}
\mathrm{Rank}(\mathbf{\Lambda}_{k,n}) \geq \mathrm{Rank}(\mu^{\rm{1}}_{n} \mathbf{I}) - \mathrm{Rank}(\tilde{\mathbf{H}}_{n}) =  N_{t}-1.
\end{align}

From \eqref{proof1_1}, it follows that $\mathrm{tr}\left(\mathbf{\Lambda}_{k,n}\mathbf{W}_{k,n}\right) = 0$, indicating that all columns of $\mathbf{W}_{k,n}$ lie in the null space of $\mathbf{\Lambda}_{k,n}$.
Since $\mathrm{rank}(\mathbf{\Lambda}_{k,n}) \geq N_{t} - 1$, the dimension of its null space satisfies $\mathrm{dim}\left(\mathrm{Null}(\mathbf{\Lambda}_{k,n})\right) \leq 1$.
Consequently, the column space of $\mathbf{W}_{k,n}$ is confined to a subspace of dimension at most one, leading to
\begin{align}
\mathrm{rank}(\mathbf{W}_{k,n}) \leq \mathrm{dim}\left(\mathrm{Null}(\mathbf{\Lambda}_{k,n})\right) \leq 1.
\end{align}
Furthermore, since $\mathbf{W}_{k,n} \succeq \mathbf{0}$ and the optimization problem admits a nontrivial feasible solution (i.e., $\mathbf{W}_{k,n} \neq \mathbf{0}$ for optimality), it follows that $\mathbf{W}_{k,n}$ must be a rank-one matrix.
Therefore, $\mathrm{rank}(\mathbf{W}_{k,n}) = 1$, completing the proof.

\end{appendices}

\bibliographystyle{ieeetr}
\bibliography{myref}
 \begin{IEEEbiography}
[{\includegraphics[width=1in,height=1.25in]{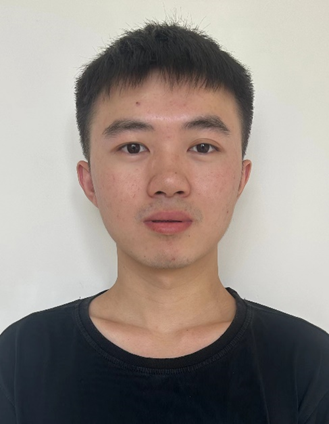}}]{Jun Wu} (Graduate Student Member, IEEE) received the B.E. degree in the School of Electronic Engineering from  Southwest Jiaotong University, China, in 2021. He is currently working toward the Ph.D. degree in the School of Automation and Intelligent  Manufacturing, Southern University of Science and Technology, Shenzhen, China. He was a visiting student at Sungkyunkwan University in 2025. His research focuses on the area of integrated sensing and communications, UAV communications, low-altitude wireless networks, and convex optimization. He was a recipient of the Best Paper Award from IEEE WCNC 2026.
\end{IEEEbiography}

 \begin{IEEEbiography}
[{\includegraphics[width=1in,height=1.25in]{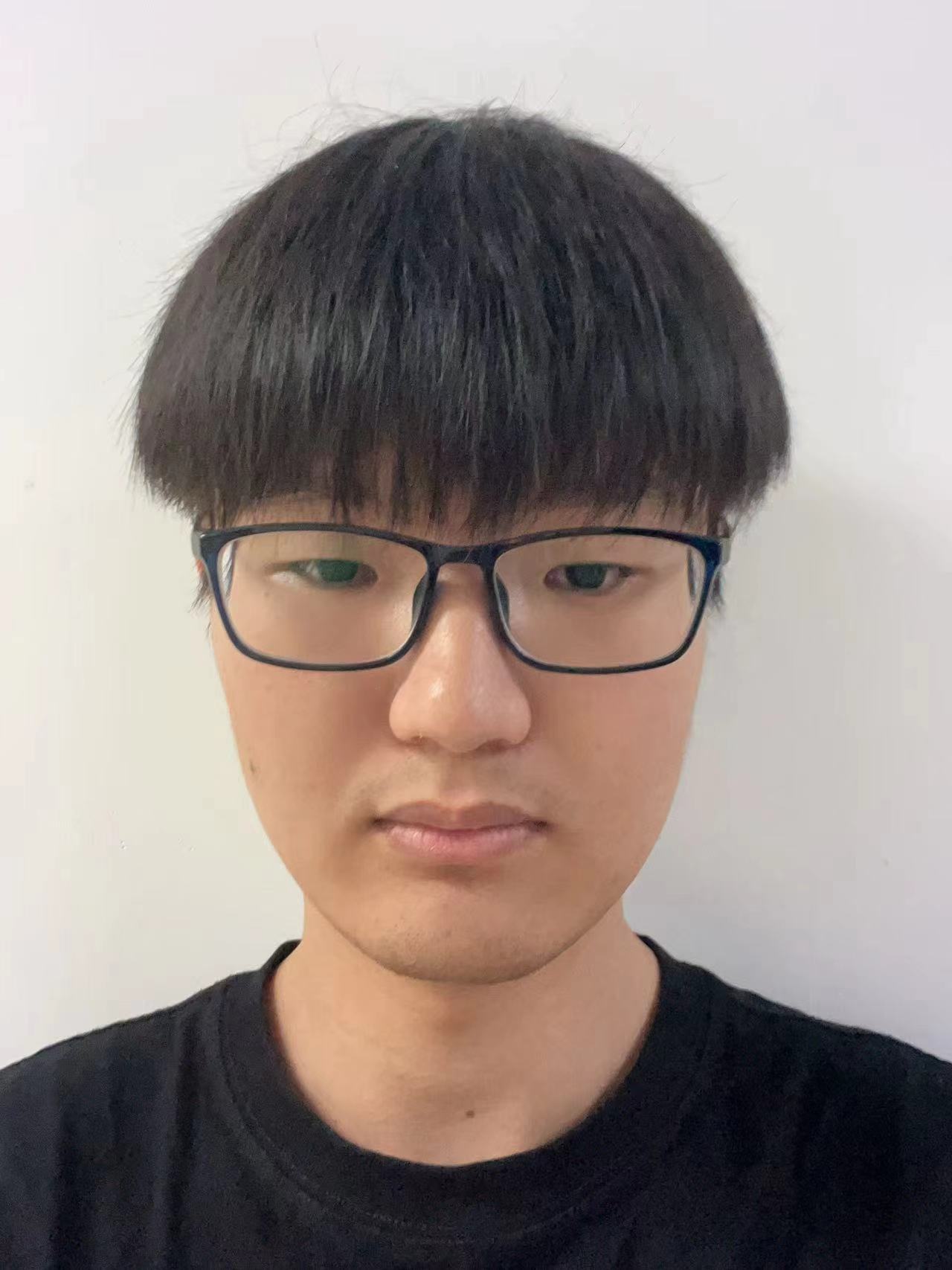}}]{Wenchao Liu}
(Graduate Student Member, IEEE) received the B.Sc. degree in Mechanical Electronics Engineering from Soochow University, Suzhou, China, in 2022, and the M.Sc. degree in Electronic Science and Technology from Southern University of Science and Technology, Shenzhen, China, in 2025. He is currently working toward the Ph.D. degree with the School of Automation and Intelligent Manufacturing, Southern University of Science and Technology, Shenzhen, China. His research interests include UAV communications, low-altitude wireless networks, and convex optimization.
\end{IEEEbiography}

\begin{IEEEbiography}[{\includegraphics[width=1in,height=1.25in]{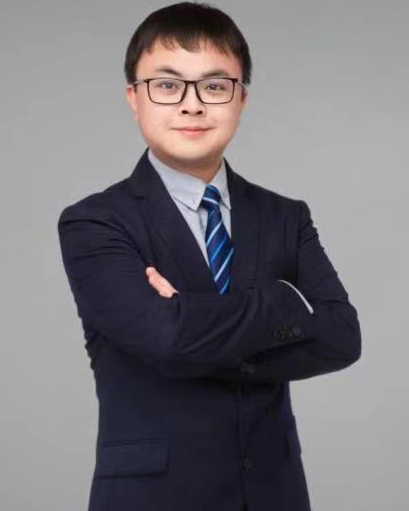}}]{Weijie Yuan} (Senior Member, IEEE) is now an Assistant Professor with the Southern University of Science and Technology. His research interests include Integrated Sensing and Communications (ISAC), Orthogonal Time Frequency Space (OTFS), and the Low-Altitude Wireless Networks (LAWN). He currently serves as an Editor for the IEEE Transactions on Wireless Communications, IEEE Transactions on Mobile Computing, IEEE Communications Magazine, IEEE Communications Standards Magazine, IEEE Transactions on Green Communications and Networking, IEEE Communications Letters, and IEEE Open Journal of Communications Society, an Guest Editor for IEEE Transactions on Vehicular Technology, IEEE Transactions on Network Science and Engineering, and IEEE Internet of Things Journal. He was the Track-Chair for IEEE ICC 2025 and IEEE VTC 2025-Spring. He served as an Organizer/the Chair of several workshops and special sessions in flagship IEEE and ACM conferences, including IEEE ICC, IEEE VTC, IEEE GlobeCom, IEEE/CIC ICCC, IEEE SPAWC, IEEE WCNC, IEEE ICASSP, and ACM MobiCom. 
He was a recipient of the Best Editor from IEEE CommL, the Best Paper Award from IEEE ICC 2023, IEEE/CIC ICCC 2023, and IEEE GlobeCom 2024, as well as the 2025 IEEE Communications Society \& Information Theory Society Joint Paper Award. 
\end{IEEEbiography}
\begin{IEEEbiography}
[{\includegraphics[width=1.2in,height=1.2in]{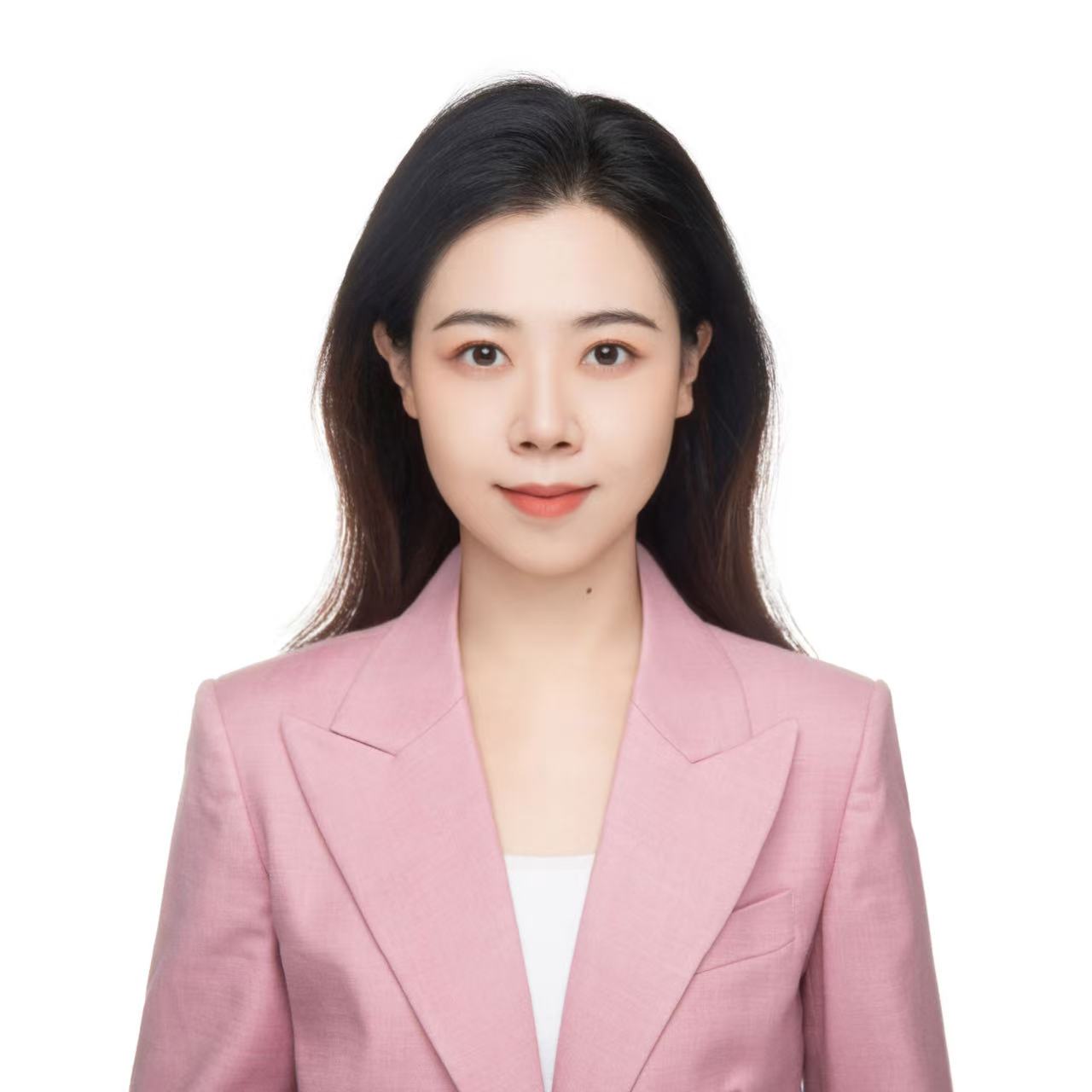}}]{Nanchi Su} (Member, IEEE) received the B.E. and M.E. degrees from the Harbin Institute of Technology, Heilongjiang, China, in 2015 and 2018, respectively, and the Ph.D. degree from University College London, London, U.K., in 2023. She is currently an Associate Professor with the Guangdong Provincial Key Laboratory of Aerospace Communication
and Networking Technology, Harbin Institute of Technology (Shenzhen), Shenzhen, China. Her research interests include integrated sensing and communication systems (ISAC), constructive interference design, physical layer security, radar signal processing, SAGIN, and situational awareness. She serves as the Lead Guest Editor of a Special Issue of the IEEE OJ-COMS and serves as workshop Co-Chair for IEEE ICC 2026, IEEE WCNC 2026 and PIMRC 2026.
\end{IEEEbiography}
\end{document}